\documentclass[twocolumn,preprintnumbers,amsmath,amssymb,prb,superscriptaddress]{revtex4-1}

\usepackage{multirow}
\usepackage{amsfonts}
\usepackage{amsopn}
\usepackage[english]{babel}
\usepackage[T1]{fontenc}
\usepackage{times}
\usepackage{mathrsfs}
\usepackage{graphicx}
\usepackage{dcolumn}
\usepackage{bm}
\usepackage[colorlinks,bookmarks=true,citecolor=blue,linkcolor=red,urlcolor=blue]{hyperref}
\usepackage[tight, nooneline]{subfigure}
\usepackage{footnote}
\usepackage{braket}

\newcommand{\bb}[1]{\mathbf{#1}}

\begin{document}
\author{B\l a\.{z}ej Jaworowski}
\affiliation{Department of Theoretical Physics, Faculty of Fundamental Problems of Technology, Wroc\l aw University of Science and Technology, Wybrze\.{z}e Wyspia\'{n}skiego 27, 50-370  Wroc\l aw, Poland}
\author{Nicolas Regnault}
\affiliation{Laboratoire Pierre Aigrain, Ecole normale sup\'erieure, PSL University, Sorbonne Universit\'e, Universit\'e Paris Diderot, Sorbonne Paris Cit\'e, CNRS, 24 rue Lhomond, 75005 Paris France}
\author{Zhao Liu}
\affiliation{Zhejiang Institute of Modern Physics, Zhejiang University, Hangzhou 310027, China}
\affiliation{Freie Universit\"at Berlin, Dahlem Center for Complex Quantum Systems and Institut f\"ur Theoretische Physik, Arnimallee 14,
14195 Berlin, Germany}

\title{Characterization of quasiholes in two-component fractional quantum Hall states and \\ fractional Chern insulators in $|C|=2$ flat bands}
\begin{abstract}
We perform an exact-diagonalization study of quasihole excitations for the two-component Halperin $(221)$ state in the lowest Landau level and for several $\nu=1/3$ bosonic fractional Chern insulators in topological flat bands with Chern number $|C|=2$. Properties including the quasihole size, charge, and braiding statistics are evaluated. For the Halperin $(221)$ model state, we observe isotropic quasiholes with a clear internal structure, and obtain the quasihole charge and statistics matching the theoretical values. Interestingly, we also extract the same quasihole size, charge, and braiding statistics for the continuum model states of $|C|=2$ fractional Chern insulators, although the latter possess a ``color-entangled'' nature that does not exist in ordinary two-component Halperin states. We also consider two real lattice models with a band having $|C|=2$. There, we find that a quasihole can exhibit much stronger oscillations of the density profile, while having the same charge and statistics as those in the continuum models.


\end{abstract}
\maketitle

\section{Introduction}
While identical particles in three spatial dimensions obey either bosonic or fermionic statistics, fractional statistics beyond these two elementary cases exists when particles are restricted in two spatial dimensions \cite{leinaas1977theory, wilczek1982quantum, frohlich1990braid}. Exotic particles with fractional statistics, known as anyons, can be realized as fractionally charged quasiparticle excitations in topologically ordered systems \cite{wen1989vacuum,wen1990topological,WenZoo}. In particular, quasiparticles with non-Abelian statistics \cite{moore1991nonabelions} are key resources for fault-tolerant quantum computation \cite{kitaev2003fault,sarma2005topologically}.

As representative topologically ordered systems, fractional quantum Hall (FQH) states\cite{tsui,laughlin} in two-dimensional (2D) electron gas penetrated by a strong magnetic field are prominent platforms to host anyons. Some experimental indications of the predicted fractional statistics of quasielectron and quasihole excitations in FQH states 
have been observed using quasiparticle interferometers\cite{kivelson1990semiclassical,chamon1997two,rosenow2012telegraph,bishara2009interferometric, bonderson2006detecting,camino2005realization,camino2007e,Willett8853}. The design of these experiments can be aided by microscopic characterization of quasiparticles, which have been done for various Abelian and non-Abelian one-component FQH states. These studies evaluate the quasiparticle size\cite{johri2014quasiholes,liu2015characterization,wu2014braiding,zaletel2012exact, toke2007nature,prodan2009mapping,storni2011localized} and simulate the braiding process\cite{tserkownyak2003monte,liu2015characterization,wan2008fractional,wu2014braiding,
zaletel2012exact,storni2011localized,barbaran2009numerical,prodan2009mapping}. However, apart from theoretical predictions\cite{girvin2007multicomponent, wen1995topological}, there were much less efforts\cite{paredes2002fermionizing,douglas2011imaging,jeon2005trial} to pursue a full microscopic characterization of quasiholes in multicomponent FQH systems with internal degrees of freedom such as spin, layer, and valley\cite{halperin1983theory,halperin1984statistics,ardonne1999new,goerbig2007analysis}.

Recently, theoretical\cite{SunNature, Neupert,Zoology,PRX,hierarchy,jaworowski2015fractional} and experimental\cite{FCIExperiment} works have shown that cousins of FQH states on the lattice, called fractional Chern insulators (FCIs)\cite{parameswaran2013fractional, liu2013review,Titusreview}, can emerge in a partially filled flat band with nonzero Chern number $C$\cite{Sun,Tang,flatkagome,flatstar,flatsqoct}. FCIs in $|C|=1$ flat bands (denoted $|C|=1$ FCIs) can be mapped to one-component FQH states to which the adiabatic continuity has been explicitly established\cite{AdiabaticFQHE,PhysRevB.87.035306,PhysRevB.86.085129}. The microscopic characterization of quasiholes in these $|C|=1$ FCIs has confirmed the correspondence to one-component FQH states in the quasiparticle level\cite{kapit2012nonabelian,nielsen2015anyon,liu2015characterization,PhysRevB.92.125105,2018arXiv180402002R}: (i) the density distribution around one quasihole in a $|C|=1$ FCI can be mapped to that in the continuum by choosing an appropriate length unit on the lattice, allowing the estimation of quasihole size on the lattice once the quasihole size in the corresponding one-component FQH state is known\cite{liu2015characterization}, and (ii) quasiholes of $|C|=1$ FCIs have the same braiding statistics as those in the corresponding one-component FQH states. However, there is an obvious lack of similar studies of quasiholes of FCIs in $|C|>1$ flat bands\cite{wang2012fractional,liu2012fractional,PhysRevB.86.241112,sterdyniak2013series,moller2015fractional,wang2013tunable}, which cannot be simply mapped to ordinary multicomponent FQH states due to their ``color-entangled'' nature\cite{wu2013bloch,wu2014haldane,PhysRevB.91.041119,PhysRevLett.116.216802}. Indeed, such FCIs can be related to multicomponent FQH systems with extended twist defects. It is unclear whether such an exotic nature causes discrepancy between the quasihole properties in $|C|>1$ flat bands and in ordinary multicomponent FQH systems.



In this work, we perform a direct characterization of quasiholes of bilayer FQH states in the continuum and FCIs in $|C|=2$ flat bands (denoted $|C|=2$ FCIs). Motivated by the relevance for the ultracold gas implementation, we focus on the $(221)$ Halperin state at $\nu_{\textrm{FQH}}=2/3$\cite{halperin1983theory,halperin1984statistics} and bosonic $|C|=2$ FCIs at $\nu_{\textrm{FCI}}=1/3$\cite{liu2012fractional,wang2012fractional,wang2013tunable,moller2015fractional,sterdyniak2013series}. With the help of numerical exact diagonalization, we calculate the particle density around a single quasihole pinned by an impurity potential, then estimate its size and charge. We also simulate the braiding process between two quasiholes by adiabatically exchanging the positions of their pinning potentials to extract the fractional statistics.
We start our study from the Halperin $(221)$ model state in Sec.~\ref{sec:221}. In this case, each quasihole must be pinned by a potential with a specific layer index. The density profile around a quasihole is isotropic and shows a clear internal structure related to the distribution of total quasihole charge among layers, which agrees with earlier results of the off-resonant light scattering from ultracold atoms in the quantum Hall regime\cite{douglas2011imaging}. We consider the total density over two layers, then accurately recover the predicted quasihole charge $-e/3$ as well as the statistical phase $(p_{\alpha})_{\textrm{an}}=\pm 2\pi/3$\cite{girvin2007multicomponent, wen1995topological,paredes2002fermionizing}. In Sec.~\ref{sec:continuum}, we consider the bilayer continuum model of $\nu_{\textrm{FCI}}=1/3$ bosonic $|C|=2$ FCIs\cite{wu2013bloch,wu2014haldane}, which has a ``color-entangled'' nature that is absent in the ordinary Halperin $(221)$ model state. In this case, we find that the potential with a layer index is also necessary for pinning a quasihole. Interestingly, although the ``color-entangled'' nature of this model affects the layer-resolved density around a quasihole extending across the boundary, the quasihole size, charge and statistics show no differences from those for the Halperin $(221)$ model state, indicating that adding ``color-entangled'' nature in ordinary bilayer FQH systems does not change the key features of quasiholes. In Sec.~\ref{sec:FCI}, we study quasiholes of $|C|=2$ FCIs in two real lattice models\cite{wang2012fractional,wang2013tunable}, both of which do not explicitly include layer information. In these cases, we find that a quasihole without an internal structure can be pinned by a layer-independent onsite potential, which is impossible in the two continuum models studied above. We obtain the same quasihole charge and statistics on the lattice as in the continuum, however, the single-quasihole particle density on the lattice displays much stronger oscillations than those for the continuum model states, suggesting the deviation of these FCIs from model states. 
In Sec.~\ref{sec:conclusions}, we summarize our results, and list some open questions for future work.

\section{Quasiholes in the Halperin $(221)$ state}\label{sec:221}
We start our study with the ordinary bilayer FQH system\cite{halperin1983theory, halperin1984statistics}. In this case, we impose periodic boundary conditions separately on each layer, such that each layer has a torus geometry. We fix the torus aspect ratio to $1$. The periodic boundary conditions require the number of magnetic flux quanta $N_{\phi}$ piercing each layer as an integer, related to the corresponding torus length $L$ by $L^2=2\pi\ell_B^2 N_\phi$, where $\ell_B$ is the magnetic length. The upper and lower layers (which we will also call colors) are indexed with $\sigma=\uparrow$ and $\downarrow$, respectively, which can also be understood as any two-component degree of freedom like the spin. We populate the system with $N_{\sigma}$ particles in each layer such that the total particle number and pseudospin is $N=N_{\uparrow}+N_{\downarrow}$ and $S_z=(N_{\uparrow}-N_{\downarrow})/2$, respectively. The total filling factor is defined as $\nu_{\textrm{FQH}}=N/N_{\phi}$. Similarly, we have the layer-resolved filling factor as $\nu_{\sigma}=N_{\sigma}/N_{\phi}$. 

In this article, we will focus on  bosons at $\nu_{\textrm{FQH}}=2/3$ with a layer-independent contact interaction
\begin{eqnarray}\label{interaction}
H_{\textrm{int}}&=&\sum_{i<j=1}^{N_\uparrow} \delta(\bb {r}_{i,\uparrow}-\bb {r}_{j,\uparrow})+\sum_{i<j=1}^{N_\downarrow} \delta(\bb {r}_{i,\downarrow}-\bb {r}_{j,\downarrow})\nonumber\\
&+&\sum_{i=1}^{N_\uparrow} \sum_{j=1}^{N_\downarrow} \delta(\bb {r}_{i,\uparrow}-\bb {r}_{j,\downarrow}),
\end{eqnarray}
where $\bb {r}_{i,\sigma}$ is the 2D position of the $i$-th boson in layer $\sigma$. Note that $H_{\textrm{int}}$ preserves $S_z$. After the projection to the lowest Landau level (LLL), both intralayer and interlayer interactions reduce to the zeroth Haldane pseudopotential\cite{haldane1983fractional, haldane1990hierarchy}. For such a system, the Halperin $(221)$ state \cite{halperin1983theory, halperin1984statistics} is the densest zero-energy ground state. As a spin singlet, it occurs in the $S_z=0$ sector ($\nu_{\uparrow,\downarrow}=1/3$) and is $3$-fold degenerate on the torus. The rest of this section is devoted to the microscopic characterization of the Halperin $(221)$ quasiholes by numerical investigation using exact diagonalization.

\subsection{A single $(221)$ Halperin quasihole}\label{ssec:oneqh221}
When we add half extra flux quanta into the system, namely $N_{\phi}=(3N+1)/2$, a single Abelian $(221)$ quasihole is nucleated. Since $N_{\phi}$ has to be an integer, we require $N$ to be odd in this case. In the energy spectrum of $H_{\textrm{int}}$, zero-energy eigenstates exist in the $2S_z=\pm 1$ sectors (one state per momentum sector for each $S_z$), which are associated with the delocalized quasihole. We pin the quasihole using a layer-dependent delta potential of strength $W$ located at position $\mathbf{w}$ in layer $\sigma_0$, i.e.,
\begin{equation}
H_{\textrm{imp}}(\mathbf{w}_{\sigma_{0}})=W\sum_{i=1}^{N_{\sigma_0}} \delta(\mathbf{r}_{i,\sigma_{0}}-\mathbf{w}_{\sigma_{0}}). 
\label{eq:deltapotential}
\end{equation}
A direct diagonalization of $H_{\textrm{int}}+H_{\textrm{imp}}(\mathbf{w}_{\sigma_{0}})$ in the LLL gives three zero-energy ground states, whose pseudospin is $2S_z=-1$ if $\sigma_0=\uparrow$ or $2S_z=1$ if $\sigma_0=\downarrow$. These three zero-energy ground states correspond to the topological degeneracy of a localized $(221)$ quasihole on the torus \cite{wen1990ground,wen1990anyons}. However, the computational cost of this direct diagonalization
is high because the impurity potential Eq.~(\ref{eq:deltapotential}) breaks the magnetic translation symmetry on the torus.
In order to increase the numerical efficiency, we assume that $W$ is small enough such that the impurity cannot mix the zero-energy manifold of $H_{\textrm{int}}$ with excited states. Then for a specific $\sigma_0$, we can first compute the zero-energy eigenstate of $H_{\textrm{int}}$ per momentum sector with the corresponding $S_z$, where we rely on the magnetic translation symmetry to reduce the Hilbert space dimension, then diagonalize the impurity potential within this zero-energy manifold. 

\begin{figure}
\centerline{\includegraphics[width=\linewidth]{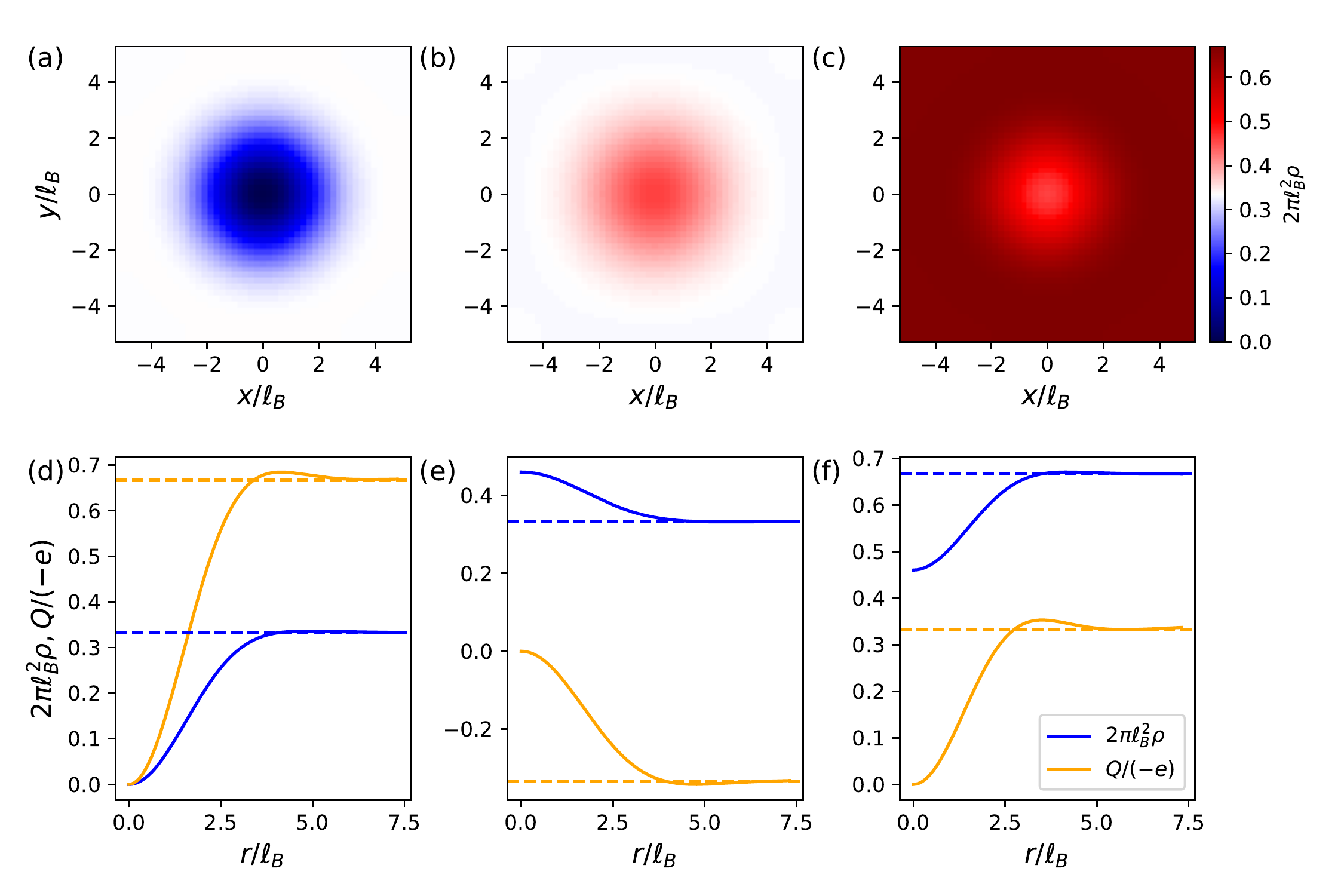}}
\caption{A single quasihole of the $(221)$ Halperin state for $N=11,N_\phi=17$, pinned by an impurity potential Eq.~(\ref{eq:deltapotential}) located in the upper layer. We show the particle density in the upper layer and lower layer and the total density in (a), (b), and (c), respectively, with the pinning potential located at the center of the plot. The corresponding radial density (blue line) and excess charge (orange line) around the quasihole are shown in (d), (e) and (f), respectively. The horizontal lines in (d), (e), (f) correspond to $2\pi \ell_B^2 \rho=1/3,Q/(-e)=2/3$ in (d), $Q/(-e)=-1/3,2\pi \ell_B^2 \rho=1/3$ in (e), and $Q/(-e)=1/3,2\pi \ell_B^2 \rho=2/3$ in (f). 
}
\label{fig:oneqh221}
\end{figure}

We focus on the particle density. In the absence of quasiholes, the particle density in each layer, after being averaged over three degenerate $(221)$ states, is uniform at $\rho_{\uparrow,\downarrow}=(1/3)/(2\pi\ell_B^2)$. We then generate a quasihole pinned by a delta potential, Eq.~(\ref{eq:deltapotential}), located in the upper layer at a point corresponding to the center of Figs.~\ref{fig:oneqh221}(a)--\ref{fig:oneqh221}(c). In this case, the obtained three zero-energy ground states of the impurity potential have $2S_z=-1$. Computing the particle density for a single ground state gives similar, however slightly state-dependent results. Such differences should disappear for large enough systems. To reduce this finite-size effect, we consider $\rho_\sigma$ and the total density $\rho_{\textrm{tot}}=\sum_\sigma \rho_\sigma$, averaged over the three ground states, as shown in Fig.~\ref{fig:oneqh221}(c) for $N=11, N_{\phi}=17$. 
Interestingly, particle densities in both layers deviate from the uniform case even though the impurity potential only acts in the upper layer, indicating an internal structure of the quasihole: $\rho_\uparrow$ drops to zero at the position of the impurity potential [Fig.~\ref{fig:oneqh221}(a)], but meanwhile $\rho_\downarrow$ develops a peak at the same position [Fig.~\ref{fig:oneqh221}(b)]; both tend to the uniform value $(1/3)/(2\pi\ell_B^2)$ when the distance $r$ from the center reaches $r \approx 4\ell_B$. Such an internal structure reflects the interlayer correlation in the $(221)$ state. We note that a similar distribution of charge among the layers in a Halperin quasihole was obtained in Ref.~\onlinecite{douglas2011imaging}. Assuming isotropic $\rho_\uparrow$ and $\rho_\downarrow$ [supported by Figs.~\ref{fig:oneqh221}(a) and \ref{fig:oneqh221}(b)], we measure the excess charge in each layer by\cite{johri2014quasiholes, liu2015characterization} 
\begin{equation}
Q_{\sigma}(r)=2\pi e \int_0^r \left[ \rho_{\sigma}(r')-\rho_{0,\sigma} \right] r' \textrm{d}r',
\label{eq:excesscharge}
\end{equation}
where $e$ is the charge of each boson, $\rho_{\sigma}(r)$ is the radial density with respect to the sample center in layer $\sigma$, and $\rho_{0,\sigma}=(1/3)/(2\pi \ell_B^2)$ is the uniform density in the absence of quasiholes. We adopt the convention of $e>0$ throughout this work, thus Eq.~(\ref{eq:excesscharge}) leads to negative quasihole charge. When evaluating Eq.~(\ref{eq:excesscharge}) in our finite systems, we choose $r$ along the diagonal direction from the sample center to the upper right corner.
The calculations show that the excess charge saturates to $-2e/3$ and $+e/3$ in the upper and lower layer, respectively [Figs.~\ref{fig:oneqh221}(d) and \ref{fig:oneqh221}(e)], agreeing with the fact that there is $2/3$ missing particle in the upper layer and $1/3$ excess particle in the lower layer compared to the exact filling $\nu_{\uparrow, \downarrow}=1/3$. 

The internal structure of the quasihole observed in the density profile motivates us to characterize the quasihole by the total particle density $\rho_{\textrm{tot}}$ over two layers [Figs.~\ref{fig:oneqh221}(c)~and~\ref{fig:oneqh221}(f)], which gives the excess charge as $-e/3$ in agreement with the theoretical prediction of the $(221)$ quasihole charge (see Ref.~\onlinecite{girvin2007multicomponent} and references therein for details, and Ref.~\onlinecite{wen1995topological} for a field-theoretical derivation). There are several ways to estimate the quasihole radius $R$\cite{johri2014quasiholes}, using for example the first or second moment of the particle density relative to that far from the quasihole. These moments can quantify the extent of the density fluctuation induced by the quasihole. Since the square root of the second moment gives a roughly $15\%$ larger estimation of $R$ than the first moment, here we adopt the former as a safer definition of the quasihole radius. Assuming isotropic density distribution, we define
\begin{equation}\label{rtot}
R=\sqrt{\frac{\int_0^{r_{\textrm{max}}} \left| \rho_{\textrm {tot}}(r)-\rho_{\textrm {tot}} (r_{\textrm{max}}) \right| r^3 \textrm{d}r}{ \int_0^{r_{\textrm{max}}}\left| \rho_{\textrm {tot}}(r)-\rho_{\textrm {tot}} (r_{\textrm{max}})  \right|r \textrm{d}r}},
\end{equation}
for finite systems, where $\rho_{\textrm {tot}}(r)$ is the total radial density with respect to the sample center, $r_{\textrm{max}}=L/\sqrt{2}$ is the largest radius available within the sample of length $L$, and we use numerical values of $\rho_{\textrm {tot}}$ along the diagonal direction from
the sample center to the upper right corner as $\rho_{\textrm {tot}}(r)$. For $N=11,N_{\phi}=17$, we obtain $R\approx 2.05 \ell_B$ (similar estimations by using the layer-resolved density $\rho_\sigma$ in Eq.~(\ref{rtot}) give $R_{\uparrow}\approx 2.06 \ell_B$ and $R_{\downarrow}\approx 2.29 \ell_B$). This can be compared with $R\approx 1.76 \ell_B$ of the $\nu_{\textrm{FQH}}=1/2$ Laughlin quasihole\cite{liu2015characterization}. 


\subsection{Braiding two $(221)$ Halperin quasiholes}\label{ssec:twoqh221}
Having characterized a single $(221)$ Halperin quasihole, let us now determine the quasihole statistics. To create two quasiholes, we need to add one extra flux quantum into the system, namely $N_{\phi}=(3N+2)/2$ which fixes $N$ to be even. These two delocalized quasiholes are associated with zero-energy eigenstates of $H_{\textrm{int}}$ in the $2S_z=-2,0,2$ sectors. The number of states in this manifold and their momenta can be deduced from the generalized Pauli principle \cite{haldane1991fractional,estienne2012spin}. We then pin the two quasiholes by two delta potentials located at $\mathbf{w}_1$ in layer $\sigma_0$ and $\mathbf{w}_2$ in layer $\sigma_0'$, respectively:
\begin{equation}
H_{\textrm{imp2}}(\mathbf{w}_{1,\sigma_{0}},\mathbf{w}_{2,\sigma_{0}'}) =H_{\textrm{imp}}(\mathbf{w}_{1,\sigma_{0}})+H_{\textrm{imp}}(\mathbf{w}_{2,\sigma_{0}'}),
\label{eq:twopotentials}
\end{equation}
where $H_{\textrm{imp}}(\mathbf{w}_{\sigma_{0}})$ is defined in Eq.~(\ref{eq:deltapotential}). The direct diagonalization of $H_{\textrm{int}}+H_{\textrm{imp2}}(\mathbf{w}_{1,\sigma_{0}},\mathbf{w}_{2,\sigma_{0}'})$ in the LLL gives three zero-energy ground states corresponding to two localized $(221)$ quasiholes. These three ground states have pseudospin $2S_z=-2,0$ and $2$ if $(\sigma_0,\sigma_0')=(\uparrow,\uparrow),(\uparrow,\downarrow)$, and $(\downarrow,\downarrow)$, respectively. Similarly to the strategy used in the single-quasihole case, we diagonalize Eq.~(\ref{eq:twopotentials}) in the zero-energy manifold of $H_{\textrm{int}}$ with a specific $S_z$ determined by $(\sigma_0,\sigma_0')$ to obtain the three zero-energy ground states with two localized $(221)$ quasiholes. 

The braiding of two quasiholes can be achieved by fixing $\mathbf{w}_1$ and varying $\mathbf{w}_2$ as $\mathbf{w}_2=\mathbf{w}_1+(r\cos \theta,-r\sin \theta)$
in Eq.~(\ref{eq:twopotentials}), where $r$ is a constant and $\theta$ changes from 0 to $2\pi$, corresponding to moving one quasihole clockwise around the other along a circle of radius $r$. 
The accumulated Berry phase during such a braiding is encoded in the eigenvalues of the unitary Berry matrix
\begin{eqnarray}\label{berry}
\mathcal{B}=\exp\Big\{-2\pi i\int_{0}^{2\pi}\gamma(\theta) d\theta\Big\},
\end{eqnarray}
where $\gamma_{\alpha\beta}(\theta)=i\langle\psi_\alpha(\theta)|\nabla_{\theta}|\psi_\beta(\theta)\rangle$ is the Berry connection matrix, and $|\psi_\alpha(\theta)\rangle$ are the three zero-energy ground states that we get by diagonalizing Eq.~(\ref{eq:twopotentials}) for each $\theta$. After choosing a smooth gauge $\braket{\psi_{\alpha}(\theta)|\psi_{\beta}(\theta+d\theta)}=\delta_{\alpha\beta}+O(d\theta^2)$ (see Ref.~\onlinecite{kapit2012nonabelian} for more details), we have $\mathcal{B}_{\alpha\beta}=\braket{\psi_{\alpha}(2\pi)|\psi_{\beta}(0)}$. The eigenvalues of $\mathcal{B}$ are $\{e^{-ip_\alpha}\}_{\alpha=1,2,3}$, where $p_\alpha$'s are the Abelian Berry phases. We choose $p_\alpha\in[0,2\pi)$ throughout the paper.

\begin{figure}
\centerline{\includegraphics[width=\linewidth]{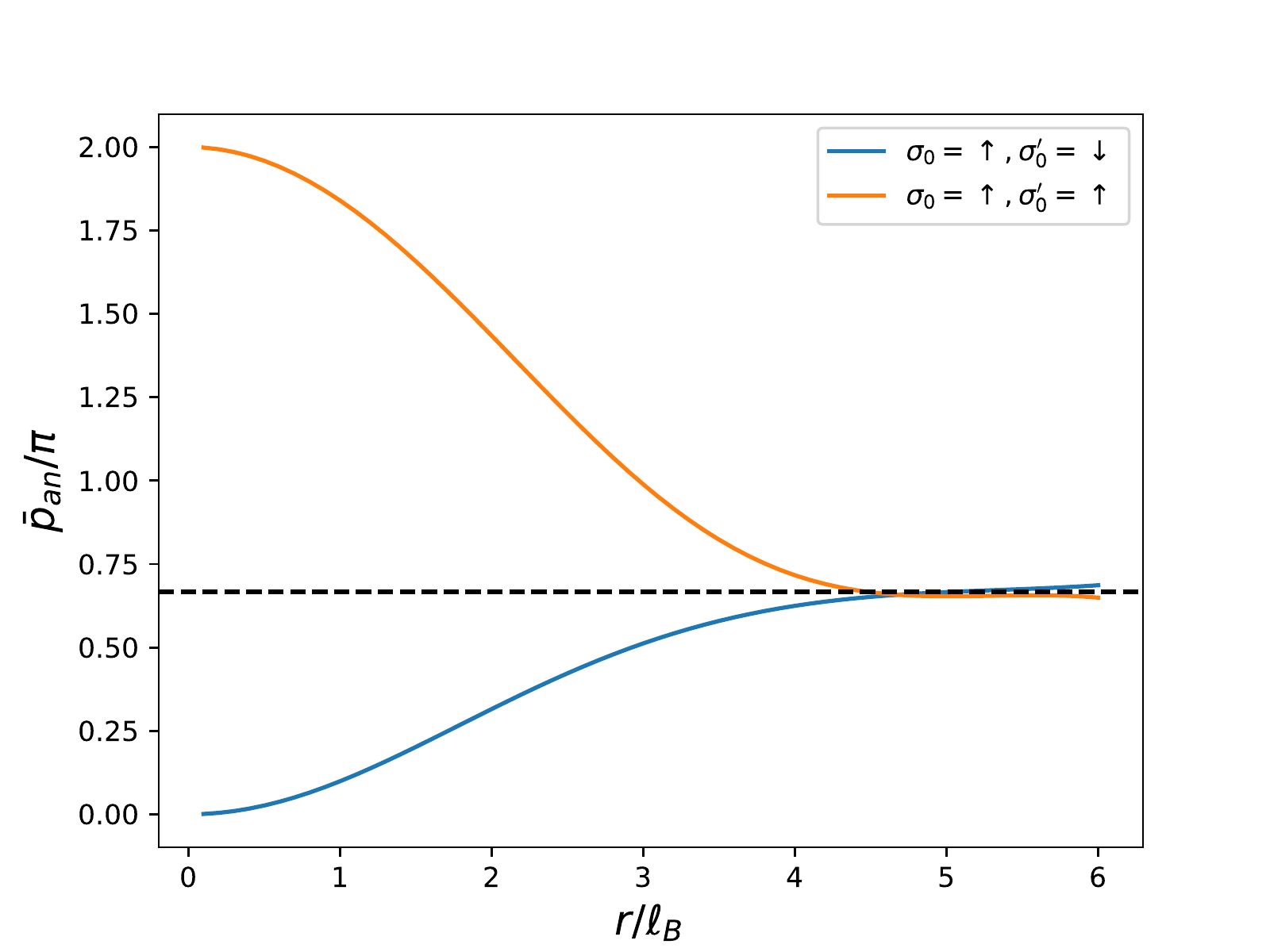}}
\caption{The anyonic phase of clockwise braiding one $(221)$ Halperin quasihole around another static one along a circular path of radius $r$. The system size is $N=10,N_{\phi}=16$. The blue (orange) curve corresponds to the case where the two impurity potentials are located in different layers (in the same layer). The dashed line corresponds to $\bar{p}_{\textrm{an}}=2\pi/3$.}
\label{fig:braiding221}
\end{figure}

The Berry phases obtained using this procedure have two origins: one is the Aharonov-Bohm (AB) phase $(p_{\alpha})_\textrm{AB}=\pi(r/\ell_B)^2/3$ caused by moving
a single quasihole in the uniform magnetic field along the same path without the other quasihole enclosed, where we have considered the total excess charge $-e/3$ over two layers; the other contribution comes from the anyonic phase $(p_\alpha)_{\textrm{an}}$. Therefore, the pure anyon statistics is $(p_{\alpha})_\textrm{an}=p_{\alpha}-\pi(r/l_B)^2/3$. We show the pure anyon statistics as a function of $r$ in Fig.~\ref{fig:braiding221}. Because all three $(p_{\alpha})_{\textrm{an}}$ are almost equal, we plot the mean $\bar{p}_{\textrm{an}}=\sum_{\alpha}(p_{\alpha})_{\textrm{an}}/3$. In the case of locating both impurity potentials in the upper layer (the orange line in Fig.~\ref{fig:braiding221}), we find that $\bar{p}_{\textrm{an}}$ converges to $2\pi/3$ for large enough $r$, being consistent with the theoretical prediction of the exchanging phase of two $(221)$ quasiholes\cite{wen1995topological,paredes2002fermionizing}. Moreover, we also recover the predicted anyon statistics at large $r$ when locating two impurity potentials in different layers (the blue line in Fig.~\ref{fig:braiding221}), which means that the density peak shown in Fig.~\ref{fig:oneqh221}(b) does behave as a part of the quasihole during the braiding, thus further confirming the internal structure of the $(221)$ quasihole. The anyon statistics deviates from the theoretical prediction at small $r$ due to the overlap between two quasiholes. The critical value of $r$ for which $\bar{p}_{\textrm{an}}$ is close enough to
$2\pi/3$ can be used as another definition of the quasihole size. One can see that $\bar{p}_{\textrm{an}}$ reaches $2\pi/3$ at $r\approx 4.5\ell_B$ in both cases, leading to $R\approx 2.25 \ell_B$.

\subsection{Layer-independent pinning potential}\label{ssec:qhpair221}
In $|C|=2$ lattice models (except those constructed by multi-orbital or layer stacking\cite{PhysRevB.86.241112,PhysRevB.91.041119,PhysRevLett.116.216802}), we will in general not have the luxury of a layer-dependent pinning potential. It is thus relevant to consider a layer-independent potential 
$\tilde{H}_{\textrm{imp}} (\mathbf{w})=\sum_{\sigma_0=\uparrow,\downarrow}H_{\textrm{imp}} (\mathbf{w}_{\sigma_0})$ first for the $(221)$ state. Doing so when we have one quasihole, i.e. $N_{\phi}=(3N+1)/2$, we obtain three ground states of $\tilde{H}_{\textrm{imp}} (\mathbf{w})$, but, in contrast to the case of $H_{\textrm{imp}} (\mathbf{w})$, they are only approximately degenerate, and they have finite energy. 

\begin{figure}
\centerline{\includegraphics[width=\linewidth]{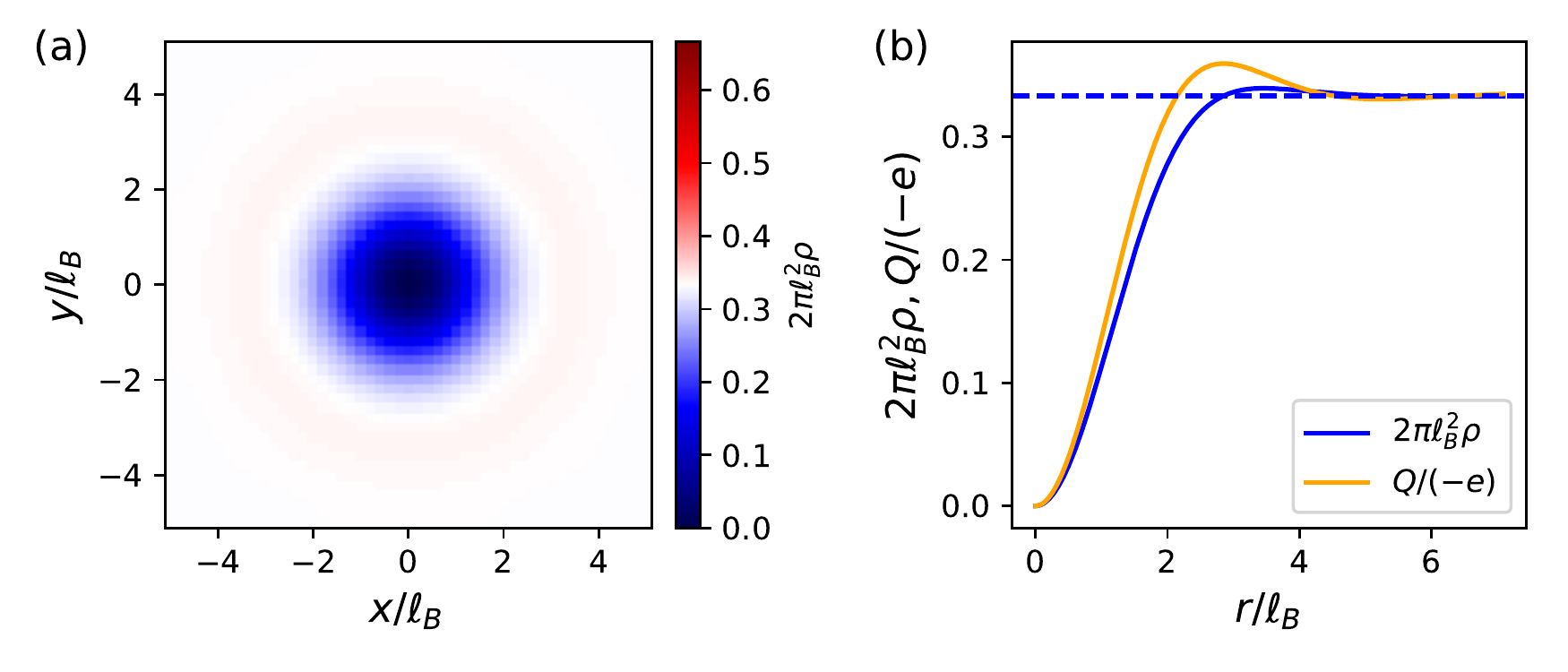}}
\caption{A pair of $(221)$ Halperin quasiholes localized on top of each other in a $N=10, N_\phi=16$ system. (a) The particle density in the upper layer, which is the same as that in the lower layer. (b) The radial density and excess charge in the upper layer. The dashed horizontal line corresponds to $2\pi \ell_B^2\rho=1/3,Q/(-e)=1/3$.}
\label{fig:qhpair221}
\end{figure}

In the situation of two quasiholes, i.e. $N_{\phi}=(3N+2)/2$, we end up in the case $S_z=0$ described in Subsection~\ref{ssec:twoqh221} with $\mathbf{w}_{1}=\mathbf{w}_{2}=\mathbf{w}$ and $(\sigma_0,\sigma_0')=(\uparrow,\downarrow)$. 
The zero-energy ground states of $\tilde{H}_{\textrm{imp}}(\mathbf{w})$ correspond to a pair of quasiholes localized on top of each other in two different layers, for which the particle density and excess charge are shown in Fig.~\ref{fig:qhpair221}. In this case, the density profile is identical in the two layers. As the density excess (depletion) of one quasihole is partially canceled by the density depletion (excess) of the other quasihole, we expect an excess charge $-e/3$ in both layers, which is exactly what we observe [Fig.~\ref{fig:qhpair221}(b)]. Interestingly, we find a reduction of the quasihole size compared with that of a single quasihole evaluated in Subsection~\ref{ssec:oneqh221}. Using Eq.~(\ref{rtot}) with the particle density per layer in Fig.~\ref{fig:qhpair221}, we estimate the quasihole radius as $R\approx 1.73 \ell_B$ when two $(221)$ quasiholes are on top of each other. Note that this value is obviously smaller than the size of a single $(221)$ quasihole, but almost the same as the $\nu_{\textrm{FQH}}=1/2$ Laughlin quasihole radius\cite{liu2015characterization}. Such a reduction of the quasihole size is a result of the interplay between two quasiholes when their pinning potentials are dragged towards each other and finally located on top of each other. 
It is also possible to braid two such pairs of quasiholes around each other. To do so, we set $N_{\phi}=(3N+4)/2$, even $N$, and pin each pair of quasiholes by a layer-independent potential $\tilde{H}_{\textrm{imp}} (\mathbf{w})$. Using the procedure similar to the one in Subsection~\ref{ssec:twoqh221}, we obtain a braiding phase close to $2 \pi /3$ for large $r$, which is consistent with previous results ($4\cdot 2 \pi /3~\textrm{mod}~2\pi =2\pi/3$).

\section{Quasiholes in the continuum analogs of $|C|=2$ FCIs}\label{sec:continuum}
In ordinary multilayer FQH systems, periodic boundary conditions are imposed separately on each layer (like in Sec.~\ref{sec:221}), i.e., the layer index of a particle does not change after it goes across the boundary and returns to the starting point. The number of magnetic flux quanta per layer, $N_\phi$, is required to be an integer in this case. Naively, such an ordinary $|C|$-layer FQH system could be thought of as the continuum analog of an FCI in a Bloch band with Chern number $C$\cite{barkeshli2012topological}, where $N_\phi$ is related to the number of lattice unit cells $N_x\times N_y$ by $N_\phi=N_x N_y/|C|$. However, this analogy meets a fundamental difficulty for $|C|>1$ when $N_x N_y$ is not divisible by $|C|$. This problem was later solved by introducing a new type of $|C|$-layer FQH systems, in which the layer index of a particle also changes when it goes across the boundary\cite{wu2013bloch}. Therefore, unlike in ordinary $|C|$-layer FQH systems, different layers (or equivalently, colors) are now connected at the boundary by an extended twist defect. In this sense, we can call these new systems as ``color-entangled'' $|C|$-layer FQH systems. The color-entangled nature removes the restriction of integer $N_\phi$ that appears in ordinary $|C|$-layer FQH systems. Indeed, by establishing the adiabatic continuity, numerical calculations have confirmed that the color-entangled $|C|$-layer FQH states are proper continuum analogs of FCIs in $|C|>1$ flat bands\cite{wu2013bloch}. 

In this section, we will first recall the basics of color-entangled $|C|$-layer FQH systems, then investigate their quasiholes for the $|C|=2$ case. The study of $|C|=2$ FCI quasiholes in real lattice models will be left in Sec.~\ref{sec:FCI}.


\subsection{Color-entangled multilayer FQH systems}\label{ssec:contmodel}

We consider a rectangular $C$-layer FQH system of dimensions $L_{x}\times L_{y}$, pierced by $N_{\phi}=L_x L_y/(2\pi \ell_B^2)$ magnetic flux quanta in each layer. Our analysis below can also be generalized to tilted systems. We assume $C>0$ in the remaining part of this section. The layers (or equivalently, the colors) are indexed by $\sigma=0,1,\dots,C-1$. The total number of orbitals in the system is $N_{s}=CN_{\phi}$, which can be factorized as $N_s=N_x N_y$. Note that the aspect ratio $L_x/L_y$ of the system does not depend on the choice of $N_x$ and $N_y$ for a fixed $N_s$. We require $C,N_s,N_x,N_y\in\mathbb{Z}$. However, unlike in ordinary $C$-layer systems, we do not restrict to integer values of $N_{\phi}$.

As shown in Ref.~\onlinecite{wu2013bloch}, a set of LLL basis states $\{|\bf k\rangle\}$ compatible with both integer and fractional $N_\phi$ can be constructed in the Brillouin zone $k_{x}=0,1,\dots, N_{x}-1$ and $k_{y}=0,1,\dots, N_{y}-1$, where $\mathbf{k}=(k_x,k_y)\in{\mathbb Z}^2$. Under the Landau gauge ${\mathbf A}\propto(0,x)$, the wavefunction of $|\bf k\rangle$ in real space ${\bf r}=(x,y)$ for color $\sigma$ is given by
\begin{multline}
\psi_{\mathbf{k}}({\bf r}_\sigma)=\braket{{\bf r}_\sigma|\mathbf{k}}\\=\frac{1}{\sqrt{\sqrt{\pi}N_{x}L_y\ell_B}}\sum_{n}^{\mathbb{Z}}e^{i2\pi(nC+\sigma)k_x/N_x} \times \\
\exp \left\{ i\frac{2\pi}{L_y} \left( k_y+n N_y +\frac{\sigma}{C}N_y \right)y \right\} \times \\
\exp \left\{
-\frac{1}{2\ell_B^2}\left[ x - \frac{2\pi \ell_B^2}{L_y }\left(k_y+nN_y+\frac{\sigma}{C}N_y  \right) \right]^2
\right\}.
\label{eq:contwfn}
\end{multline} 
One can verify that $\psi_{\mathbf{k}}({\bf r}_\sigma)$ obeys boundary conditions
\begin{eqnarray}
\label{cebc}
T_x(L_x)P^{N_x}\psi_{\mathbf{k}}({\bf r}_\sigma)&=&\psi_{\mathbf{k}}({\bf r}_\sigma),\nonumber\\
T_y(L_y)Q^{N_y}\psi_{\mathbf{k}}({\bf r}_\sigma)&=&\psi_{\mathbf{k}}({\bf r}_\sigma),
\end{eqnarray}
where $T_{x}$ (resp. $T_{y}$) is the magnetic translation operator in the $x$ (resp. $y$) direction acting on the coordinate ${\bf r}$, and $P$ and $Q$ are color operators acting on the color index $\sigma$ as
\begin{eqnarray}
P\ket{\sigma}&=&\ket{(\sigma +1) ~\textrm{mod}~C},\nonumber\\
Q\ket{\sigma}&=&e^{2\pi i \sigma/C}\ket{\sigma}.
\end{eqnarray}
When $N_x$ is divisible by $C$ (implying that $N_\phi$ is an integer), as $P^{N_x}$ is an identity, the boundary conditions in Eq.~(\ref{cebc}) return to those of ordinary $C$-layer FQH systems (up to a layer-dependent flux insertion induced by $Q$). However, when $N_{x}$ is not divisible by $C$ (fractional $N_\phi$), the simultaneous appearance of magnetic translation and color operators in the boundary conditions entangles the color degrees of freedom at the boundary. Indeed, in this case, the layer index of a particle is shifted at the boundary according to $T_x(L_x)\psi({\bf r}_\sigma)=P^{-N_x}\psi({\bf r}_\sigma)=\psi({\bf r}_{(\sigma-N_x)~\textrm{mod}~C})$, i.e., different layers are connected by an extended twist defect at the boundary. Such a color-entangled nature is absent in ordinary $C$-layer FQH systems. Note that the specific dependence of the color-entangled nature on $N_x$ is due to a gauge choice in $\psi_{\mathbf{k}}({\bf r}_\sigma)$. Indeed, it would depend on $N_y$ for the other Landau gauge ${\mathbf A}\propto(-y,0)$.

The single-particle basis $\{|\bf k\rangle\}$ leads to a uniform Berry curvature in the Brillouin zone and Chern number $C$\cite{wu2013bloch}, thus mimicking a Bloch band with Chern number $C$ on a lattice of $N_x\times N_y$ unit cells. Similarly to the situation of ordinary FQH systems, color-entangled FQH model states can be constructed. Cousins of $C$-layer Halperin states can be defined as the zero-energy ground states of suitable Haldane's pseudopotentials diagonalized under the basis $\{|\bf k\rangle\}$ at a specific filling $\nu=N/(N_xN_y)$ \cite{wu2013bloch}. These zero modes can be regarded as the continuum analogs of lattice FCIs in Chern number $C$ bands\cite{wu2013bloch}. In the following, we will investigate the quasiholes in the $C=2$ case for bosons interacting via the color-independent zeroth Haldane's pseudopotential Eq.~(\ref{interaction}). This interaction Hamiltonian, after being diagonalized under the basis $\{|\bf k\rangle\}$, gives three degenerate zero-energy ground states for any $N,N_s,N_x$ and $N_y$ at $\nu=1/3$, which are the continuum analogs of numerically observed $C=2$ FCIs at $\nu_{\textrm{FCI}}=1/3$\cite{wang2012fractional,liu2012fractional,PhysRevB.86.241112,sterdyniak2013series,moller2015fractional,wang2013tunable}.

\subsection{Quasiholes of color-entangled $\nu=1/3$ FQH states}\label{ssec:oneqhcont}

We choose $L_x=L_y=L$ without any loss of generality in our numerical calculations. We first generate a single quasihole in the color-entangled $\nu=1/3$ FQH state by adding one extra orbital into the system, i.e., putting $N_s=3N+1$. For even $N_s$, because either $N_x$ or $N_y$ must be even, we can always make the color-entangled nature trivial by choosing a suitable Landau gauge. Therefore, we focus on odd $N_s$, in which case $N_\phi$ must be fractional and the $C=2$ color-entangled FQH systems is truly different from the ordinary bilayer ones. To make the comparison with Sec.~\ref{sec:221} easier, we will use the notation $\sigma=\uparrow$ or $\downarrow$ instead of $\sigma=0$ or $1$. The delocalized quasihole is associated to a zero-energy manifold of the interaction Hamiltonian $H_{\textrm{int}}$ [Eq.~(\ref{interaction})], containing one state per momentum sector. We again use a layer-resolved, i.e., a color-resolved delta potential of strength $W$ located at position $\mathbf{w}$ in layer $\sigma_0$, i.e.,
$W\sum_{i=1}^{N} \delta(\mathbf{r}_{i,\sigma_{0}}-\mathbf{w}_{\sigma_{0}})$ to pin the quasihole. Diagonalizing the impurity potential in the zero-energy manifold of $H_{\textrm{int}}$ gives three zero-energy ground states for each value of $\sigma_0$, corresponding to a localized quasihole.

\begin{figure}
\centerline{\includegraphics[width=\linewidth]{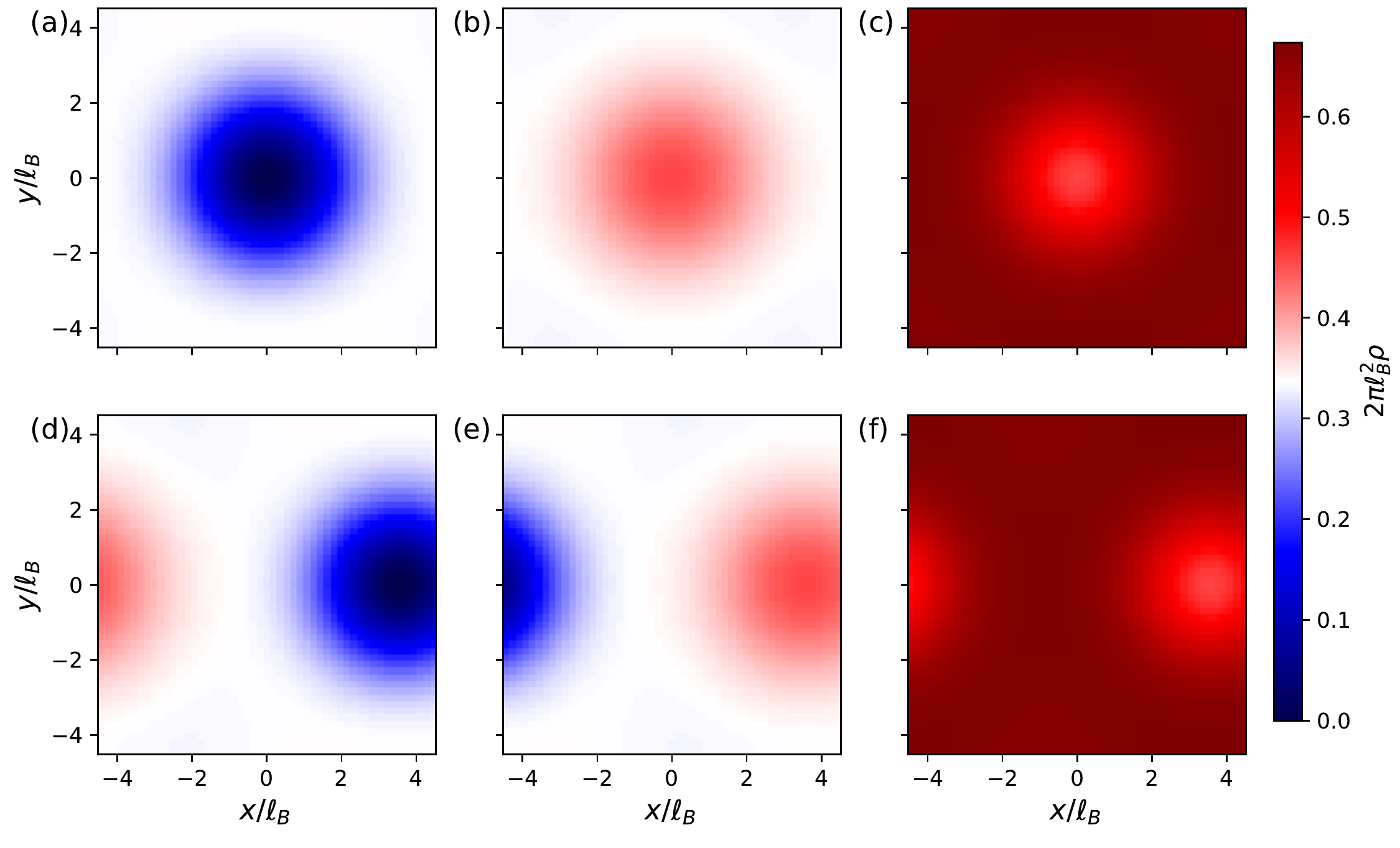}}
\caption{A single quasihole of the color-entangled $\nu=1/3$ FQH state for $N=8,N_s=25,N_x=N_y=5$, pinned by a delta potential in the upper layer. The extended twist defect connecting two layers is located at $x=\pm L/2$. In (a), (b), and (c), we show the particle density in the upper layer and lower layer and the total density, respectively, with the pinning potential located at ${\bb w}=(0,0)$. In (d), (e), and (f), the same quantities are shown for the pinning potential located at ${\bb w}=(2L/5,0)$.
}
\label{fig:snapshots1}
\end{figure}

In Fig.~\ref{fig:snapshots1}, we show the particle density around the quasihole for $N=8,N_s=25,N_x=N_y=5$ and $\sigma_0=\uparrow$. In order to make the effect of the color-entangled nature explicit, we arbitrarily assume that the extended twist defect connecting two layers is located at $x=L/2$ [equivalent to $x=-L/2$ due to the boundary conditions Eq.~(\ref{cebc})]. Similarly to what we saw in Sec. \ref{ssec:oneqh221}, the quasihole has an internal structure. When the quasihole is far from the twist defect, the layer-resolved particle densities $\rho_{\uparrow,\downarrow}$ [Figs.~\ref{fig:snapshots1}(a) and \ref{fig:snapshots1}(b)] are almost identical to those in the ordinary $(221)$ state shown in Figs.~\ref{fig:oneqh221}(a) and \ref{fig:oneqh221}(b). This is as expected because the quasihole does not feel the defect. On the contrary, when the quasihole is pinned near the twist defect, the density depletion and excess change their layer indices as the quasihole extends across the twist defect [Figs.~\ref{fig:snapshots1}(d) and \ref{fig:snapshots1}(e)], reflecting the effect of color-entangled boundary conditions Eq.~(\ref{cebc}). However, the total density over two layers with respect to the quasihole is identical, irrespective of the quasihole position [Figs.~\ref{fig:snapshots1}(c) and \ref{fig:snapshots1}(f)]. Therefore, we have the same quasihole size in both cases, which is also the same as that of the ordinary $(221)$ state.

\begin{figure}
\centerline{\includegraphics[width=\linewidth]{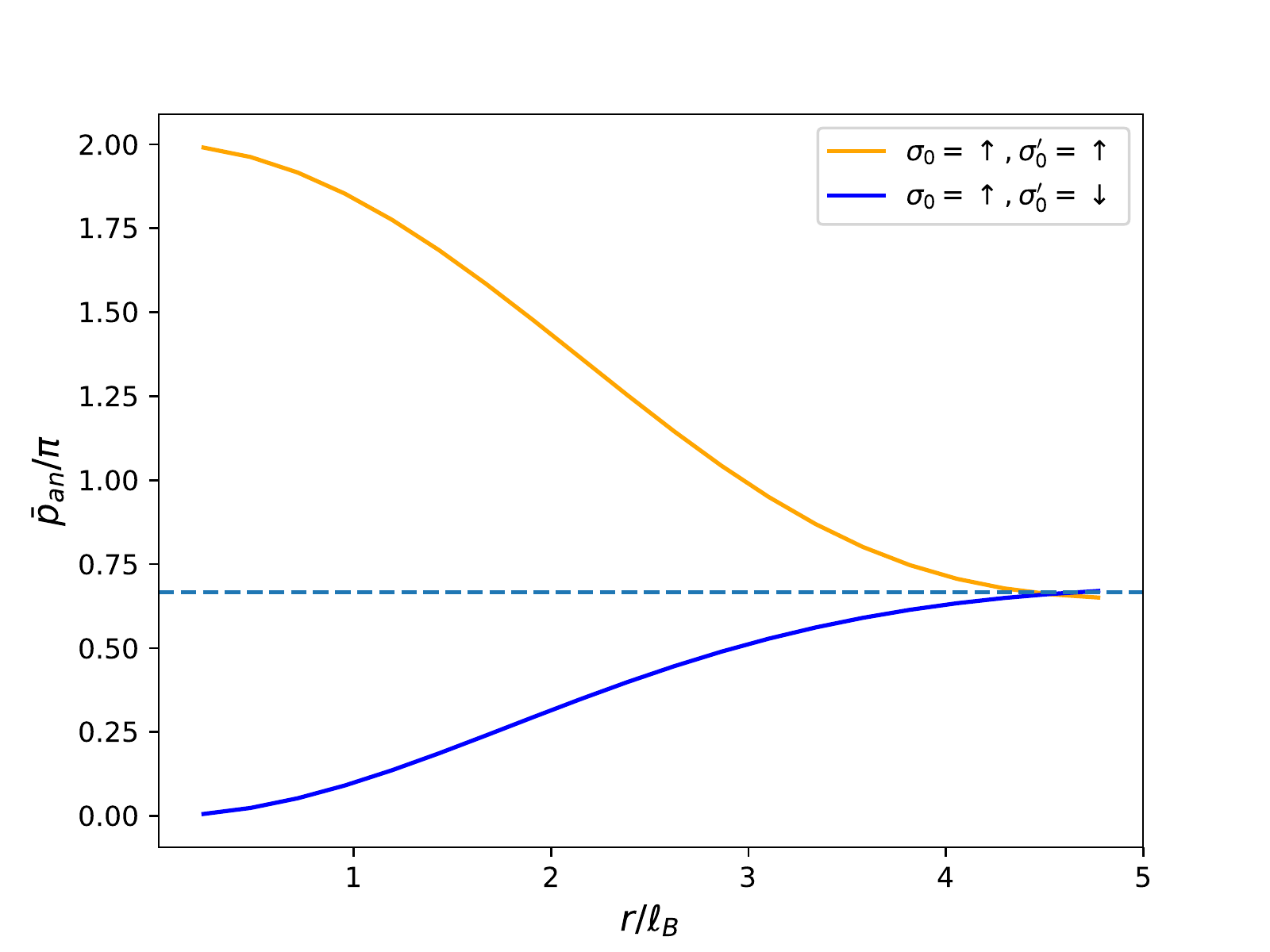}}
\caption{The anyonic phase of clockwise braiding one quasihole around another static one along a circular path of radius $r$ for the color-entangled $\nu=1/3$ FQH state. The system size is $N=9,N_s=29,N_x=1,N_y=29$. The aspect ratio of the sample is determined by $L_x/L_y=1$ rather than $N_x/N_y$. The braiding path crosses with the twist defect at $r\approx L/10\approx 0.95\ell_B$, as described in the text. The blue (orange) curve corresponds to the case where the two impurity potentials are initially located in different layers (in the same layer). The dashed line corresponds to $\bar{p}_{\textrm{an}}=2\pi/3$.
}
\label{fig:snapshots2}
\end{figure}

We then set $N_s=3N+2$ to generate two quasiholes of the color-entangled $\nu=1/3$ FQH state. These two delocalized quasiholes are associated with the zero-energy eigenstates of $H_{\textrm{int}}$, the number and momentum of which can be deduced from the generalized Pauli principle \cite{wu2014haldane}. In order to braid the two quasiholes, we pin one of them by a delta potential located at $\mathbf{w}_1$ in layer $\sigma_0$, i.e., $W\sum_{i=1}^{N} \delta(\mathbf{r}_{i,\sigma_{0}}-\mathbf{w}_{1,\sigma_{0}})$, where $\mathbf{w}_1$ and $\sigma_0$ is fixed. The other quasihole is dragged around the static one along a clockwise circular path of radius $r$ by a mobile potential $W\sum_{i=1}^{N} \delta(\mathbf{r}_{i,\sigma_{0}'}-\mathbf{w}_{2,\sigma_{0}'})$ located at $\mathbf{w}_2$ in layer $\sigma_0'$, where $\mathbf{w}_2=\mathbf{w}_1+(r\cos \theta,-r\sin \theta)$. Diagonalizing the sum of two impurity potentials in the zero-energy manifold of $H_{\textrm{int}}$ gives three zero-energy ground states for any $(\sigma_0,\sigma_0')$, corresponding to two localized quasiholes. Again we assume that the extended twist defect connecting two layers is located at $x=\pm L/2$. The layer index $\sigma_0'$ of the mobile impurity potential will flip when the braiding path crosses with the defect. In order to probe the interplay between the defect and braiding properties, we put the static quasihole near $x=L/2$. In Fig.~\ref{fig:snapshots2}, we show the anyonic braiding phase obtained by the same method as that used in Sec.~\ref{ssec:twoqh221}, as a function of $r$ for $\mathbf{w}_1=(2L/5,0)$ and different initial $(\sigma_0,\sigma_0')$. For this setting, the braiding path starts to cross with the defect at $x=L/2$ when $r\approx L/10$. We do not observe any discontinuity of the braiding phase near this point. In fact, the results at all $r$'s are almost the same as those for the ordinary $(221)$ state (Fig.~\ref{fig:braiding221}). As expected for any twist defect without endpoints\cite{barkeshli2014symmetry}, the color-entangled nature does not affect the quasihole braiding statistics when $|C|=2$. 

\section{Quasiholes in $|C|=2$ FCIs on the lattice}\label{sec:FCI}
\subsection{Lattice models}\label{ssec:FCImodels}
\begin{figure}
\centerline{\includegraphics[width=\linewidth]{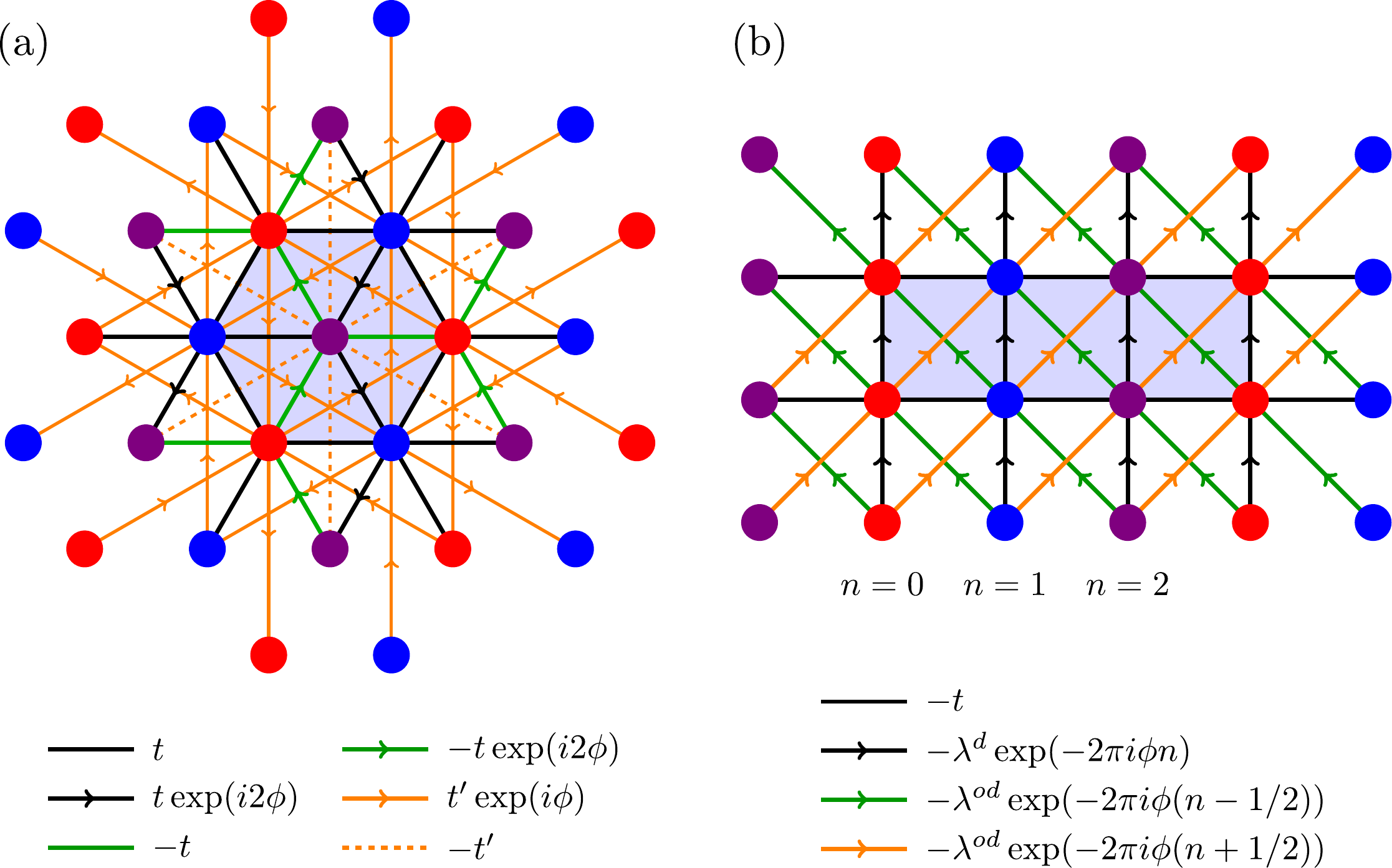}}
\caption{ The hopping terms in (a) the triangular lattice model and (b) the generalized Hofstadter model. The unit cell in each model is covered by a gray polygon. We show hoppings starting or ending within this unit cell. The hopping coefficients are given below the lattice configuration of each model. For a complex hopping, the given phase in the hopping coefficient is obtained when a particle moves along the arrow direction. In (b), the value of $n$ in a hopping coefficient is the $x$ position of the starting site of the hopping.  }
\label{fig:lattices}
\end{figure}

Having investigated the quasiholes in ordinary and color-entangled bilayer FQH states, let us now move to lattice quasiholes in $|C|=2$ FCIs. We study two lattice models: the triangular lattice model \cite{wang2012fractional} and the generalized Hofstadter model on a square lattice \cite{wang2013tunable}. Particles can hop between nearest-neighboring and next-nearest-neighboring sites in both models. The hopping terms that define the respective tight-binding models are given in Fig.~\ref{fig:lattices}. We adopt parameters $t=1$, $t'=1/4$, $\phi=\pi/3$ for the triangular lattice model [Fig.~\ref{fig:lattices}(a)] and $t=1, \lambda^{\textrm{d}} = 1, \lambda^{\textrm{od}} = -1/2, \phi=1/3$ for the generalized Hofstadter model [Fig.~\ref{fig:lattices}(b)], such that the lowest band of each model carries $|C|=2$. 


We then consider $N$ bosons on a finite lattice of $N_{s}$ unit cells under periodic boundary conditions. Since there are $N_{s}$ single-particle states in the lowest band, the band filling factor is defined as $\nu_{\textrm{FCI}}=N/N_{s}$. 
For bosons interacting via the onsite repulsion
\begin{equation}
\label{intlat}
H_{\textrm{int}}^{\textrm{lat}}=U\sum_{i}n_i(n_i-1),
\end{equation}
where $n_{i}$ is the boson number operator on lattice site $i$ and $U$ is the interaction strength. We will use the flat band approximation, meaning that we set the band gap to infinity first (projection onto the lowest band), and then take $U$ to be large, neglecting the band dispersion. In this approximation, $U$ is the sole energy scale which can be set to $U=1$. Both lattice models have three quasi-degenerate gapped ground states at $\nu_{\textrm{FCI}}=1/3$, corresponding to $|C|=2$ FCIs\cite{wang2012fractional, wang2013tunable}. In this section, we evaluate the quasihole properties of these FCIs using exact diagonalization. Moreover, considering the correspondence between $|C|=2$ FCIs and color-entangled bilayer FQH states, we study both even and odd $N_s$ to expose the possible color-entangled effect in the lattice systems.

\subsection{Tilted lattice}\label{ssec:tilted}

Before presenting results on FCI quasiholes, let us first elaborate on our choice of lattice samples. The most common choices are regular samples spanned by the two primitive lattice vectors $\mathbf{a}_{1}$ and $\mathbf{a}_{2}$, with $N_{i}$ unit cells in the direction of $\mathbf{a}_{i}$ and $N_1 N_2 =N_s$. However, for some $N_s$'s, we cannot find suitable factors $N_1$ and $N_2$ such that the system aspect ratio is close to $1$. When the system size is too small in one direction, the FCI phase has an instability towards CDW \cite{bernevig2012thin}. This problem especially plagues the generalized Hofstadter model with $\mathbf{a}_{1}=(3,0)$ and $\mathbf{a}_{2}=(0,1)$, whose unit cell is very elongated in one direction. 


In order to overcome this difficulty, we also consider tilted samples which can keep the lattice aspect ratio close to $1$\cite{repellin2014z2, hierarchy}. 
In that case, the sample is a parallelogram spanned by vectors $\mathbf{T}_{1}=n_{1,1}\mathbf{a}_{1}+n_{1,2}\mathbf{a}_{2}$ and $\mathbf{T}_{2}=n_{2,1}\mathbf{a}_{1}+n_{2,2}\mathbf{a}_{2}$, where $n_{1,1},n_{1,2},n_{2,1},n_{2,2} \in \mathbb{Z}$. It contains $N_s=\frac{\|\mathbf{T}_1\times \mathbf{T}_2\|}{\|\mathbf{a}_1\times \mathbf{a}_2\|}=|n_{1,1}n_{2,2}-n_{1,2}n_{2,1}|$ unit cells. The regular cases just correspond to the special choice of $n_{1,1}=N_1,n_{1,2}=0,n_{2,1}=0,n_{2,2}=N_2$. We then define the lattice aspect ratio for a tilted sample as
\begin{equation}
A=\begin{cases}
\|\mathbf{T}_2-\textrm{proj}_{\mathbf{T}_1}(\mathbf{T}_2)\|/\|\mathbf{T}_1\| & \textrm{if} ~ \|\mathbf{T}_1\|  \geq \|\mathbf{T}_2\|, \\
\|\mathbf{T}_1-\textrm{proj}_{\mathbf{T}_2}(\mathbf{T}_1)\|/\|\mathbf{T}_2\| & \textrm{if} ~ \|\mathbf{T}_1\|  < \|\mathbf{T}_2\|
\end{cases},
\label{eq:lataratio}
\end{equation}
where $\textrm{proj}_{\mathbf{A}}(\mathbf{B})$ denotes the projection of vector $\mathbf{B}$ onto the direction of vector $\mathbf{A}$. For a given $N_{s}$, we can choose suitable $n_{1,1},n_{1,2},n_{2,1},n_{2,2}$ to keep $A$ as close to $1$ as possible. 

Under periodic boundary conditions in the directions of $\mathbf{T}_1$ and $\mathbf{T}_{2}$, we can construct the first Brillouin zone containing $N_s$ momentum points. 
These momenta are used in our numerics to do band projection and label many-body eigenstates. We refer to Ref.~\onlinecite{repellin2014z2} for the details of extracting all allowed momenta in the first Brillouin zone of tilted samples.




\subsection{A single quasihole of $|C|=2$ FCIs at $\nu_{\textrm{FCI}}=1/3$ }\label{ssec:oneqhFCI}

Let us now study a single quasihole in $|C|=2$ FCIs. To create it, we enlarge the lattice size to $N_{s}=3N+1$. Similar to the FQH case, this quasihole is associated with a low-energy (but generally not zero-energy) manifold, containing one state per momentum sector, in the energy spectrum of $P_{\mathrm{LB}}H_{\textrm{int}}^{\textrm{lat}}P_{\mathrm{LB}}$, where $P_{\mathrm{LB}}$ is the operator of projection onto the lowest band. We apply a simple onsite impurity potential $V_{\textrm{imp}}^{\textrm{lat}}(i)=Wn_i$ to pin the quasihole on lattice site $i$. Note that this potential is layer-independent in contrast to the ones used in FQH systems, as we are considering lattice models without layer index. The situation is thus similar to Sec.~\ref{ssec:qhpair221}. Again, we assume the impurity strength $W$ cannot mix the quasihole manifold with higher-energy states. Therefore, we can first compute the quasihole manifold of $P_{\mathrm{LB}}H_{\textrm{int}}^{\textrm{lat}}P_{\mathrm{LB}}$, where we rely on the lattice translation symmetry to reduce the Hilbert space dimension, then safely diagonalize $V_{\textrm{imp}}^{\textrm{lat}}$ in this manifold to obtain the ground states with a localized quasihole. These ground states are almost three-fold degenerate.


\begin{figure}
\centerline{\includegraphics[width=\linewidth]{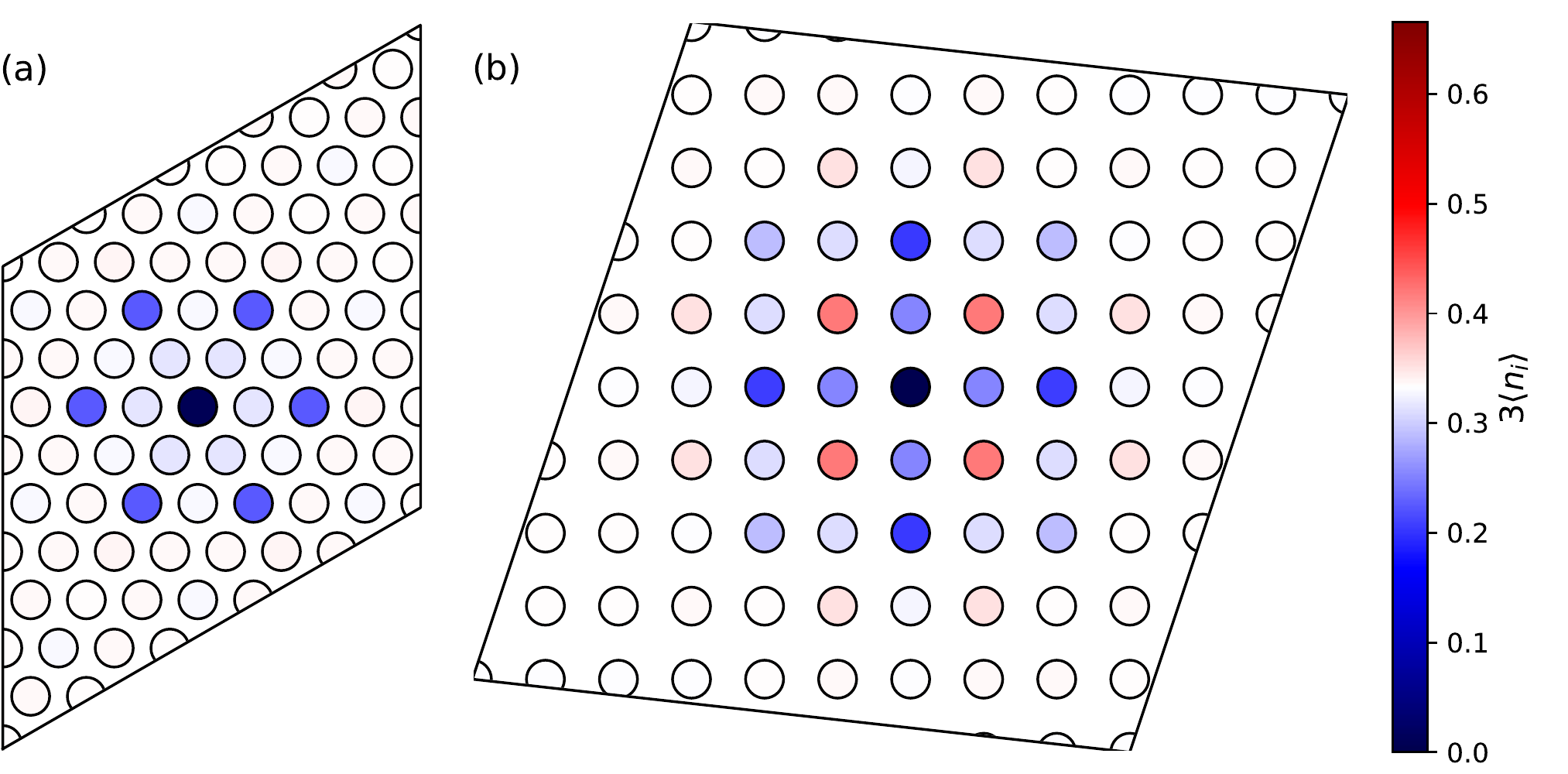}}
\caption{The lattice site occupation around one $|C|=2$ FCI quasihole at $\nu_{\textrm{FCI}}=1/3$ for (a) the triangular lattice model with $N=8$ bosons on a regular $N_1\times N_2=5\times 5$ sample; and (b) the generalized Hofstadter model with $N=9$ bosons on a tilted sample. The tilted sample is generated by $n_{1,1}=1,n_{1,2}=9, n_{2,1}=-3,n_{2,2}=1$, with $N_{s}=28$ unit cells.
}
\label{fig:oneqhFCI}
\end{figure}

\begin{figure}
\centerline{\includegraphics[width=\linewidth]{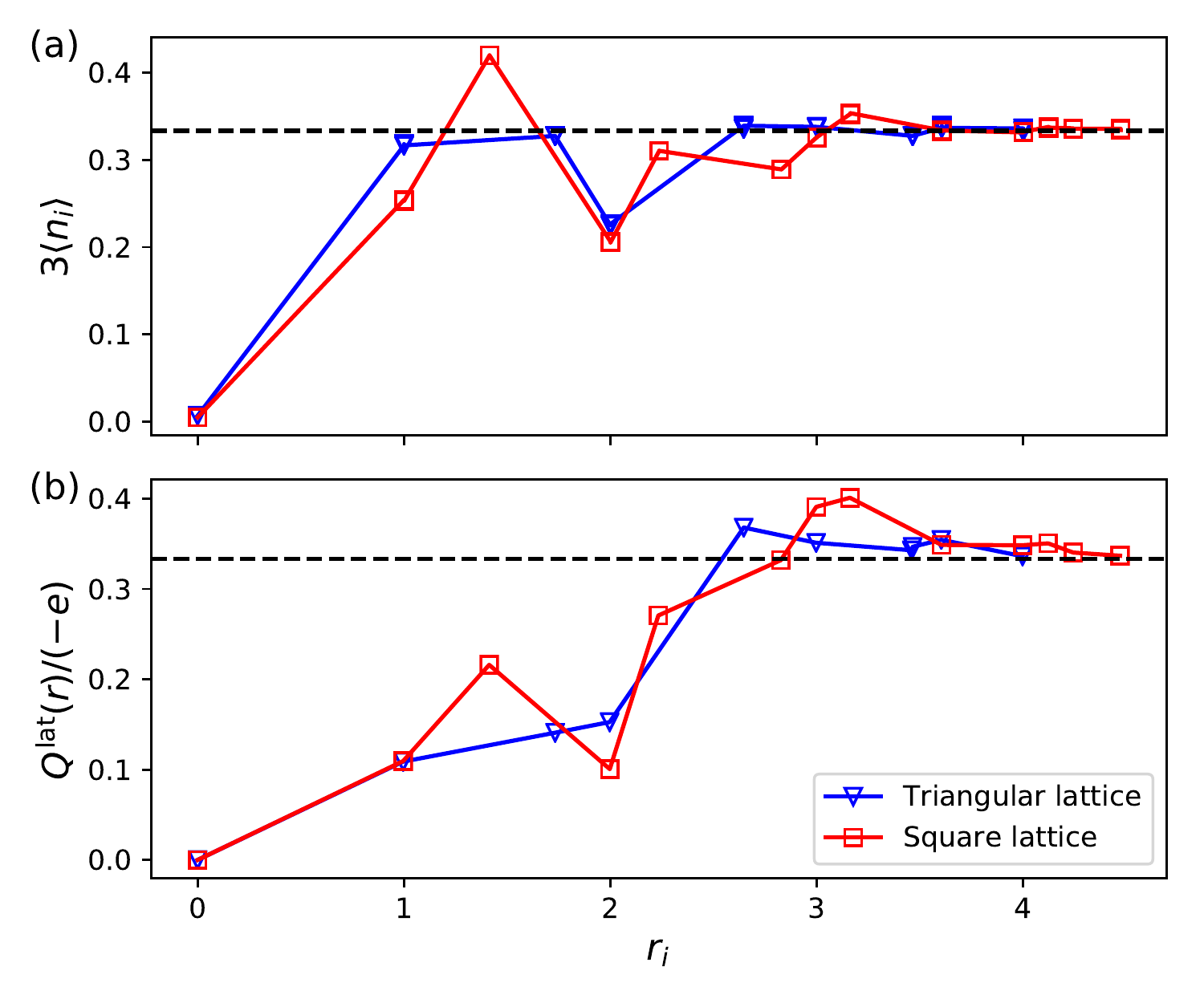}}
\caption{(a) The radial lattice site occupation and (b) the excess charge around a $|C|=2$ FCI quasihole at $\nu_{\textrm{FCI}}=1/3$, as a function of the distance $r$ (in units of the lattice constant) from the quasihole. We use the same lattices as those in Fig.~\ref{fig:oneqhFCI}. The dashed line is located at $3\langle n_i \rangle=1/3$ in (a) and $Q^{\textrm{lat}}/(-e)=1/3$ in (b).
}
\label{fig:radialdensity_FCI}
\end{figure}

We focus on the lattice site occupation $\langle n_i \rangle$. In the absence of quasiholes, we find that $3\langle n_i \rangle$ is approximately uniform at $1/3$ for each individual ground state of a finite system, where the factor $3$ is because there are three lattice sites per unit cell in both models that we consider. The deviation of $\langle n_i \rangle$ from the uniform value is a finite-size effect and can be reduced by averaging over the three degenerate ground states. In the presence of one quasihole, we thus demonstrate $\langle n_i \rangle$ averaged over the three ground states of the impurity potential for the triangular lattice model and the generalized Hofstadter model, with odd and even $N_s$, respectively (Fig.~\ref{fig:oneqhFCI}). In both cases, the quasihole is indeed pinned at the position of the impurity potential where particles are almost fully screened, leading to a very small $\langle n_i \rangle\sim 10^{-3}$. Note that this is very different from the situation of a single quasihole in bilayer FQH states, where only particles in one of the two layers are fully screened, with an particle excess in the other layer (Figs.~\ref{fig:oneqh221} and \ref{fig:snapshots1}). The density profile is closer to a Laughlin quasihole \cite{johri2014quasiholes,liu2015characterization} with a vanishing density at the center. The lattice site occupation is inversion symmetric with respect to the quasihole, and $3\langle n_i \rangle$ tends to the uniform value $1/3$ on sites far from the quasihole. However, beside the global minima at the position of the impurity potential, $\langle n_i \rangle$ also develops several local minima and maxima around the quasihole in both models, showing much stronger oscillations than the FQH density for model states studied in Secs.~\ref{sec:221} and \ref{sec:continuum}. Such oscillations can be clearly seen in the radial plot of $\langle n_i \rangle$ (Fig.~\ref{fig:radialdensity_FCI}). The excess charge on the lattice can be measured by 
\begin{equation}
Q^{\textrm{lat}}(r_i)=e\sum_{r_j<r_i} (\langle n_j \rangle - \nu_{\textrm{FCI}} /3),
\end{equation}
where $r_i$ is the distance between the site $i$ and the position of the pinning potential, and the summation goes over all sites $j$ with $r_j<r_i$. The $1/3$ factor comes from the number of sites per unit cell, i.e., $\nu_{\textrm{FCI}} /3$ is the average site occupation for the uniform distribution. 
In this way, we count the total excess of particle number compared with an FCI without quasiholes. 
We find that $Q^{\textrm{lat}}(r)$ also strongly oscillates with $r$ (Fig.~\ref{fig:radialdensity_FCI}).
Nevertheless, it approaches $-e/3$ for sufficiently large $r$, indicating that the quasihole charge of $|C|=2$ FCIs at $\nu_{\textrm{FCI}}=1/3$ is the same as that in ordinary $\nu_{\textrm{FQH}}=2/3$ and color-entangled $\nu_{\textrm{FQH}}=1/3$ FQH states. 



We can further confirm the quasihole charge from the Aharonov-Bohm phase that a quasihole picks up when moving around a unit cell. As discussed in Ref.~\onlinecite{liu2015characterization}, this phase should be equal to $\pm 2\pi q_{\textrm{qh}}/(-e)$ due to the existence of an effective magnetic field in the unit cell, where $q_{\textrm{qh}}$ is the quasihole charge. Therefore, we expect a Berry phase $\pm 2\pi/3$ for $|C|=2$ FCI quasiholes at $\nu_{\textrm{FCI}}=1/3$, where the sign depends on the direction of the path enclosing a unit cell area.


We adopt the method in Refs.~\onlinecite{kapit2012nonabelian, liu2015characterization} to move the quasihole by a time-dependent impurity. At each time $t$, this impurity potential has the form of $(1-\eta)V_{\textrm{imp}}^{\textrm{lat}}(j)+\eta V_{\textrm{imp}}^{\textrm{lat}}(k)$, which is nonzero only on two nearest-neighboring sites $j$ and $k$. When $\eta$ slowly changes with $t$ from $0$ to $1$, a quasihole is gradually moved from $j$ to $k$. Then similarly it can be moved from $k$ to the next site. We suppose the quasihole returns to the initial site at $t=T$. Similar to Sec.~\ref{sec:221}, the unitary Berry matrix is $\mathcal{B}=Pe^{-2\pi i\int_{0}^T\gamma(t) dt}$,
where $\gamma_{\alpha\beta}(t)=i\langle\psi_\alpha(t)|\nabla_{t}|\psi_\beta(t)\rangle$ is the Berry connection matrix, $|\psi_\alpha(t)\rangle$ are the approximately degenerate states that we get by diagonalizing the impurity potential at each $t$ in the quasihole manifold of $P_{\mathrm{LB}}H_{\textrm{int}}^{\textrm{lat}}P_{\mathrm{LB}}$, and $P$ is the time ordering symbol. We have checked that the three-fold degeneracy of $|\psi_\alpha(t)\rangle$ holds at any $t$. By imposing a smooth gauge condition $\langle\psi_\alpha(t)|\psi_\beta(t+dt)\rangle=\delta_{\alpha\beta}+\mathcal{O}(dt^2)$, we have $\mathcal{B}_{\alpha\beta}=\langle\psi_\alpha(T)|\psi_\beta(0)\rangle$. The eigenvalues of $\mathcal{B}$ are $\{e^{-ip_\alpha}\}_{\alpha=1,2,3}$, where $p_\alpha$'s are the AB phases that the quasihole picks up. In Table~\ref{tab:oneqhFCI}, we show the results for the largest investigated systems with either even or odd $N_s$, where we move the quasihole anti-clockwise around a unit cell indicated in Fig.~\ref{fig:lattices}. The AB phases are indeed very close to the expected $2\pi/3$ for all systems, which is consistent with the quasihole charge $-e/3$.

\begin{table*}
\caption{\label{tab:oneqhFCI}The AB phases $p_1$, $p_2$ and $p_3$ obtained by moving a single quasihole anticlockwise around a unit cell indicated in Fig.~\ref{fig:lattices}. The columns $n_{1,1}$, $n_{1,2}$, $n_{2,1}$, $n_{2,2}$ are the parameters of the tilted lattices that we have considered (as defined in Sec. \ref{ssec:tilted}). The aspect ratio $A$ is defined in Eq.~(\ref{eq:lataratio}). Note that the AB phases deviate more from $2\pi/3$ for small lattice aspect ratio $A$.}
\renewcommand{\multirowsetup}{\centering}
\begin{ruledtabular}
\begin{tabular}{cccccccccccccccc}
FCI model & $N$ & $N_s$ & $n_{1,1}$ & $n_{1,2}$ & $n_{2,1}$ & $n_{2,2}$ & $A$ & $(p_1,p_2,p_3)_{\textrm{AB}}/\pi$\\
\hline
\multirow{3}{2cm}{generalized Hofstadter}& $8$ & $25$ & $1$ & $8$ & $-3$ &  $1$ & $0.915$ & $(0.668,0.666,0.666)$\\
                                         & $9$ & $28$ & $1$ & $9$ & $-3$ &  $1$ & $0.933$ & $(0.663,0.669,0.668)$\\
                                         & $9$ & $28$ & $4$ & $0$ & $0$ &  $7$ & $0.583$ & $(0.639,0.680,0.681)$\\
\hline
\multirow{2}{2cm}{triangular}& $8$ & $25$ & $5$ & $0$ & $0$ & $5$ & $0.866$ & $(0.670,0.666,0.664)$\\
                             & $9$ & $28$ & $4$ & $0$ & $0$ & $7$ & $0.495$ & $(0.617,0.696,0.687)$\\
\end{tabular}
\end{ruledtabular}
\end{table*}

\subsection{Braiding two quasiholes of $|C|=2$ FCIs at $\nu_{\textrm{FCI}}=1/3$}\label{ssec:twoqhFCI}
We now investigate the statistics of $|C|=2$ FCI quasiholes at $\nu_{\textrm{FCI}}=1/3$. We generate two quasiholes on the lattice by considering $N_s=3N+2$. Again, these two quasiholes are associated with a low-energy manifold in the energy spectrum of $H_{\textrm{int}}^{\textrm{lat}}$, whose counting can be deduced from the generalized Pauli principle \cite{wu2014haldane}. We exchange two quasiholes or move one around the other by a time-dependent impurity similar to the one used in the last subsection. At each time $t$, this impurity potential is nonzero only on three lattice sites $j,k$ and $l$, with the form of $V_{\textrm{imp}}^{\textrm{lat}}(j)+(1-\eta)V_{\textrm{imp}}^{\textrm{lat}}(k)+\eta V_{\textrm{imp}}^{\textrm{lat}}(l)$. Here we choose $k$ and $l$ to be two nearest-neighboring sites. Such an impurity potential pins one quasihole at site $j$, and move the other one from site $k$ to site $l$ when $\eta$ varies continuously from $0$ to $1$. We then use the same method as that in the last subsection to calculate the Berry phase. 

\begin{figure}
\centerline{\includegraphics[width=\linewidth]{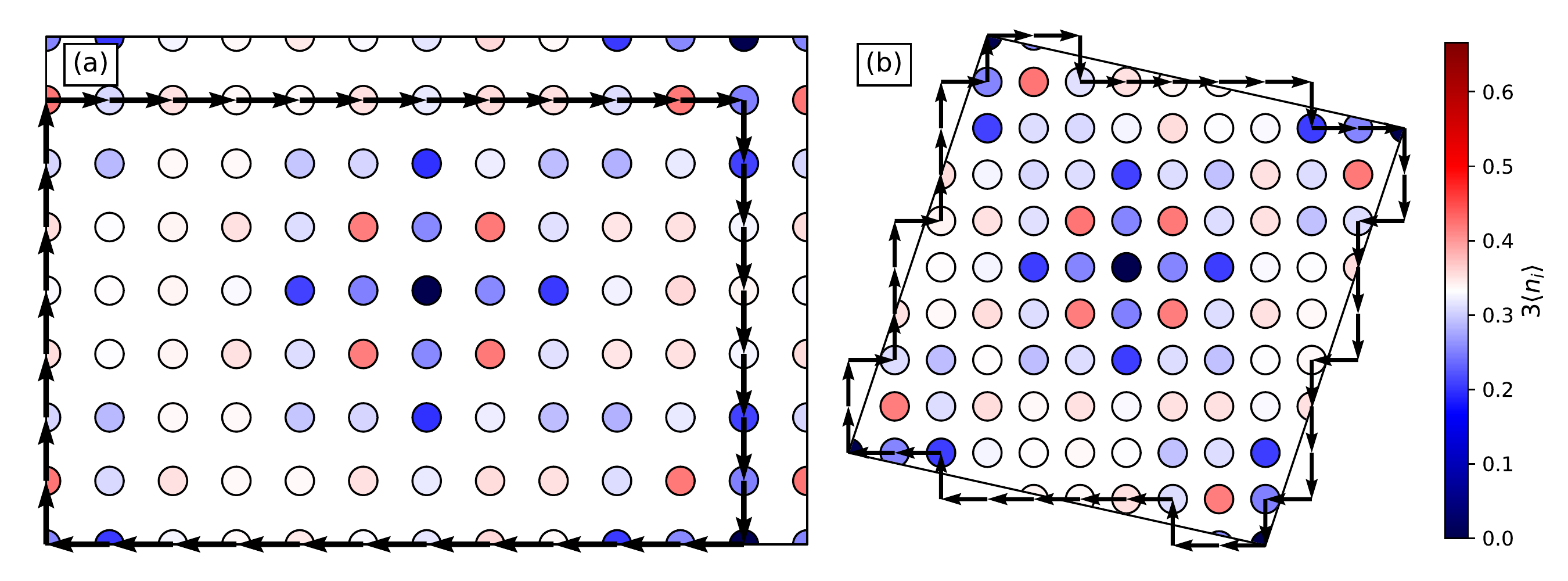}}
\caption{Braiding one quasihole around another for $|C|=2$ FCIs at $\nu_{\textrm{FCI}}=1/3$ in the generalized Hofstadter model. Here we give the braiding path on (a) a regular sample with $n_{1,1}=4,n_{1,2}=0, n_{2,1}=0,n_{2,2}=8$ and (b) a tilted sample with $n_{1,1}=1,n_{1,2}=9, n_{2,1}=-3,n_{2,2}=2$. The mobile quasihole is dragged around the static quasihole along the path indicated by the arrows, starting from the lower right corner of the sample. When the path goes outside the sample, we use periodic boundary conditions to shift the pinning potentials back into the sample. The plot of the path is superimposed on the initial lattice site occupation before the braiding.}
\label{fig:braidFCI}
\end{figure}

We focus on the generalized Hofstadter model in the following. Let us first pin one quasihole at the sample center, then move the other quasihole clockwise around it. For regular samples, the mobile quasihole simply goes through the outermost sites of the sample [Fig.~\ref{fig:braidFCI}(a)]. However, for tilted samples we choose polylines as the braiding path, connecting only the nearest-neighboring sites near the boundary [Fig.~\ref{fig:braidFCI}(b)]. This path goes slightly out of the periodic boundaries, but it reduces the error on the Berry phases. Again, the total Berry phase $p_\alpha$ can be split into two parts: one is the AB phase $(p_\alpha)_{\textrm{AB}}$ caused by moving a single quasihole along the same path without
other quasiholes enclosed; and the other is the anyonic braiding phase $(p_\alpha)_{\textrm{br}}$.
In Table~\ref{tab:braidFCI}, we show $(p_\alpha)_{\textrm{br}}=p_\alpha-(p_\alpha)_{\textrm{AB}}$ for three representative systems. Although exhibiting some distortions related to the overlap of two quasiholes during the braiding, $(p_\alpha)_{\textrm{br}}$ are always close to $2\pi/3$ for both even and odd $N_s$. Therefore, $|C|=2$ FCIs at $\nu_{\textrm{FCI}}=1/3$ have the same anyon statistics as the continuum $(221)$ and color-entangled $\nu=1/3$ FQH states.


\begin{figure}
\centerline{\includegraphics[width=\linewidth]{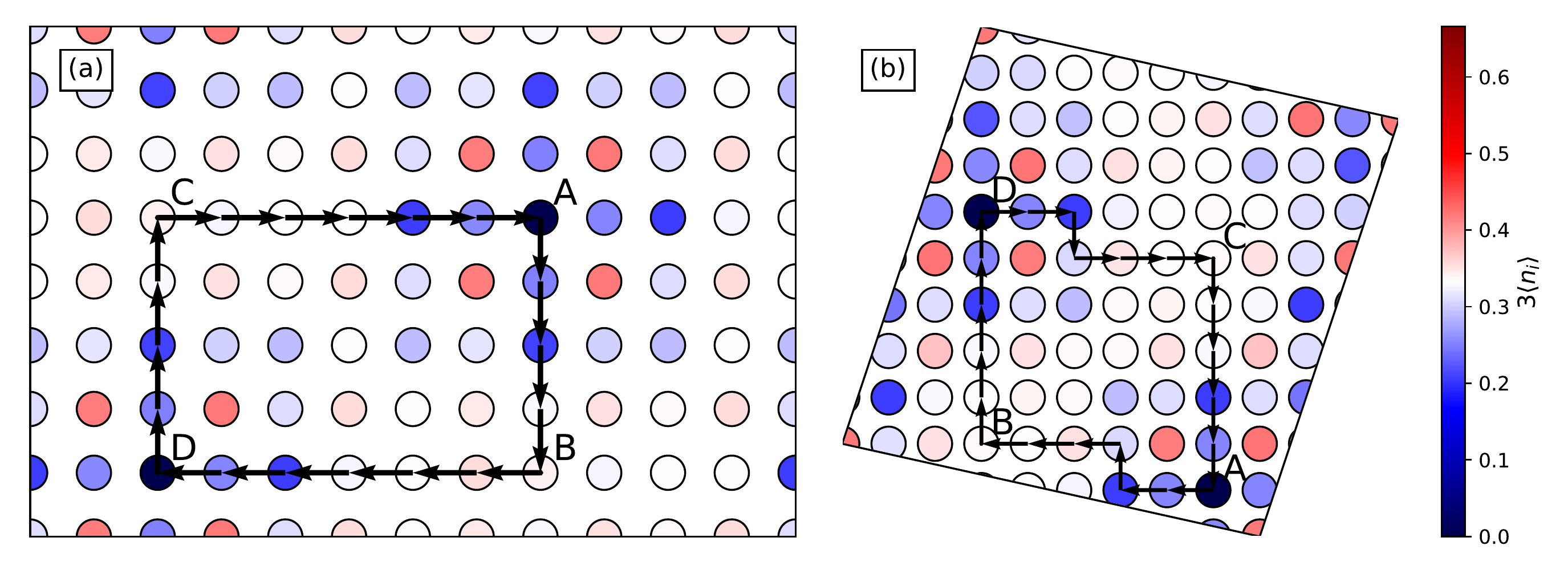}}
\caption{Exchanging two $|C|=2$ FCI quasiholes at $\nu_{\textrm{FCI}}=1/3$ in the generalized Hofstadter model. Here we give the braiding path on (a) a regular sample with $n_{1,1}=4,n_{1,2}=0, n_{2,1}=0,n_{2,2}=8$ and (b) a tilted sample with $n_{1,1}=1,n_{1,2}=9, n_{2,1}=-3,n_{2,2}=2$. The two quasiholes are moved following $(AD)\rightarrow (BD) \rightarrow (BC) \rightarrow (DC) \rightarrow (DA)$, as indicated by the arrows. The plot of the path is superimposed on the initial lattice site occupation before the exchange.}
\label{fig:excFCI}
\end{figure}

\begin{table*}
\caption{\label{tab:braidFCI} The anyon statistics for $|C|=2$ FCIs at $\nu_{\textrm{FCI}}=1/3$ in the generalized Hofstadter model. The columns $n_{1,1}$, $n_{1,2}$, $n_{2,1}$, $n_{2,2}$ and $A$ provide the geometry of the tilted lattice. The $(p_1,p_2,p_3)_{\textrm{br}}/\pi$ column contains statistics obtained by moving one quasihole around another along a clockwise path. Then, in the last column we show the statistics obtained by exchanging two quasiholes along a clockwise path. Considering the uniform magnetic field in the model, we simply approximate the AB phase as $(p_\alpha)_{\textrm{AB}}=-(8/9)\pi S$\cite{note}, where $S$ is the area enclosed by the path. 
}
\begin{center}
\begin{ruledtabular}
\begin{tabular}{cccccccccccccc}
$N$ & $N_s$ & $n_{1,1}$ & $n_{1,2}$ & $n_{2,1}$ & $n_{2,2}$ & $A$ & $(p_1,p_2,p_3)_{\textrm{br}}/\pi$ & $(p_1,p_2,p_3)_{\textrm{ex}}/\pi$ \\ \hline
$8$ &  $26$ &  $2$ &  $6$ &  $-2$ &  $7$ &  $0.918$ & $(0.690,0.631,0.651)$ & $(1.360,1.346,1.319)$ \\ 
$9$ &  $29$ &  $1$ &  $9$ &  $-3$ &  $2$ &  $0.967$ & $(0.621,0.692,0.675)$ & $(1.290,1.345,1.334)$ \\
$10$ & $32$ &  $4$ &  $0$ &  $0$ &  $8$ &  $0.666$ & $(0.686,0.619,0.649)$ & $(1.333,1.333,1.333)$ 
\end{tabular}
\end{ruledtabular}
\end{center}
\end{table*}

Further examination of the quasihole statistics can be performed by exchanging the positions of two quasiholes. The typical paths of this kind are shown in Fig.~\ref{fig:excFCI}. Initially, we pin the first and the second quasihole at points $A$ and $D$, respectively. We denote this configuration as $(AD)$. Then, we exchange two quasiholes in such a way that the configuration evolves as $(AD)\rightarrow (BD) \rightarrow (BC) \rightarrow (DC) \rightarrow (DA)$. In each of these four steps, only one quasihole is moved, and the other is static. We carefully choose the path connecting $A,B,C,D$ (especially on tilted lattices) to reduce the overlap between two quasiholes as much as possible. 
The results are also given in Table~\ref{tab:braidFCI}, where we obtain anyonic exchange phases $(p_\alpha)_{\textrm{ex}}=p_\alpha-(p_\alpha)_{\textrm{AB}}$ all close to $4\pi/3$. Note that the anyonic statistics of moving one quasihole around the other is indeed twice of that of exchanging two quasiholes (up to modulo $2\pi$). 

\section{Summary and conclusions}\label{sec:conclusions}

In this work, we perform an extensive numerical study of quasiholes in the $\nu_{\textrm{FQH}}=2/3$ bilayer Halperin $(221)$ state, the $\nu=1/3$ color-entangled bilayer FQH state, and the $\nu_{\textrm{FCI}}=1/3$ fractional Chern insulators in $C=2$ bands. For the $(221)$ model state, we pin a $-e/3$ quasihole by a layer-dependent delta impurity. The quasihole shows an internal structure with density depletion and excess among two layers, reflecting the interlayer correlation of the $(221)$ state. We use the second moment of the total particle density relative to the that far from the quasihole to measure the quasihole radius and get $R\approx 2.05\ell_B$, which is larger than the quasihole radius $R\approx 1.76\ell_B$ of the $\nu=1/2$ Laughlin quasihole\cite{liu2015characterization}. In the presence of two quasiholes, we accurately reproduce the predicted braiding phase $2\pi/3$ when two quasiholes are well separated. Interestingly, when two quasiholes are dragged close and finally located on top of each other in different layers, we find that their radius reduces to $R\approx 1.73\ell_B$ due to the interplay between them. 

Similarly, we also pin a $-e/3$ quasihole in the $\nu=1/3$ color-entangled bilayer FQH model state (with an odd total number of orbitals) by a layer-dependent delta impurity. In this case, an extended twist defect exists in the sample, connecting two layers. While the particle density in two respective layers is exchanged once the quasihole extends across the twist defect, the total density over two layers with respect to the quasihole keeps identical irrespective of the quasihole position. The  second moment of the total relative particle density gives the same quasihole radius as for the $(221)$ model state. Moreover, even if the braiding path goes across the twist defect, we still get the same braiding statistics as for the ordinary $(221)$ state, as expected\cite{barkeshli2014symmetry}. While our work focuses on the bilayer case, we argue that adding color-entangled feature in a general multicomponent FQH system does not change key quasihole properties such as the radius and statistics. 

For the two $|C|=2$ FCI models that we study, we observe the same quasihole charge and statistics as those in ordinary and color-entangled bilayer FQH states. This is true for both even and odd number of unit cells, thus the effect of the color-entangled nature on key quasihole properties is also absent on the lattice. 
In contrast to the continuum case, the two lattice models that we consider do not have a layer structure. We find that an onsite impurity potential, not referring to any internal (color) degree of freedom, is sufficient to pin a quasihole in these systems. The resulting lattice quasihole thus does not have an internal structure similar to that in the continuum, so we cannot map the density distribution on the lattice to that in the continuum by a simple rescaling of the length unit on the lattice, like what was done in Ref.~\onlinecite{liu2015characterization}. However, this may not be true for all the existing FCI models, as we expect that a layer-dependent impurity is still needed to pin a quasihole in other $|C|=2$ models constructed by layer or orbital stacking\cite{PhysRevB.86.241112,PhysRevB.91.041119,PhysRevLett.116.216802}. We also notice that the site occupation around a quasihole in the two $|C|=2$ FCI models present strong oscillations. On the contrary, such oscillations do not exist in all $|C|=1$ FCI models studied in Ref.~\onlinecite{liu2015characterization}. This difference suggests that those $|C|=1$ FCIs are closer to model states after the FCI-FQH mapping\cite{PhysRevB.86.085129,wu2013bloch} than our $|C|=2$ FCIs. 

There are several possible future developments based on this work. So far we only consider Abelian anyons in the $|C|=2$ case. It is more challenging to generalize our results to non-Abelian anyons or the $|C|>2$ case. It would be also interesting to characterize the quasiholes from other viewpoints, for example, by modular matrices extracted from the minimally entangled states\cite{zhangyi,wei1,wei2}, in which the information of braiding statistics is encoded. 
In this work, we have probed the interplay between the quasihole properties and the color-entangled nature represented by an extended twist defect without ending points. Since a microscopic model of twist defects with ending points was proposed for FCIs in Ref.~\onlinecite{liugenon}, it would be instructive to study the effect of these non-extended defects on the braiding statistics of quasiholes.



\begin{acknowledgements}
B.J. acknowledges the funding from National Science Centre (NCN), Poland, grant PRELUDIUM No. 2016/21/N/ST3/00843. N.R. was supported by the grant ANR TNSTRONG No. ANR-16-CE30-0025. Z.L. is supported by the National Thousand-Young-Talents Program of China. Z.L. was additionally supported by Alexander von Humboldt Research Fellowship for Postdoctoral Researchers. 
\end{acknowledgements}
\bibliography{quasiholes.bib}

\begin{thebibliography}{85}
\bibitem [{not()}]{note}%
  \BibitemOpen
  \href@noop {} {\bibinfo  {journal} {We find that the AB phase of moving one
  quasihole clockwise around an elementary plaquette of the generalized
  Hofstadter model is $-8\pi/9$}\ }\BibitemShut {NoStop}%
\end{thebibliography}

\end{document}